\documentclass{article}
\pdfoutput=1

\usepackage{mathtools}
\usepackage{vmargin}
\setpapersize{USletter}
\setmarginsrb{1in}{1in}{1in}{1in}{0pt}{0pt}{0pt}{12pt}

\usepackage{amsmath}
\usepackage{amsthm}
\usepackage{latexsym}
\usepackage{amssymb}
\usepackage{bm}
\usepackage{graphics}
\usepackage{url}
\usepackage{derivative}
\usepackage{graphicx}

\usepackage{enumitem}

\usepackage[vlined,ruled,boxed,linesnumbered]{algorithm2e}
\DontPrintSemicolon

\usepackage[frozencache,cachedir=.]{minted} 

\newcommand{\realring}{\mathbb{R}}

\newcommand{\tarski}{\textsc{Tarski}}

\newtheorem{definition}{Definition}

\newtheorem{theorem}{Theorem}
\newtheorem{lemma}{Lemma}
\newtheorem{corollary}{Corollary}
\newtheorem{example}{Example}

\newcommand{\true}{\textsc{true}}
\newcommand{\false}{\textsc{false}}
\newcommand{\fail}{\textsc{fail}}

\newcommand{\relop}{\sigma}

\newcommand{\mydiv}[2]{{\mathrm{div}}{\left(#1,#2\right)}}
\newcommand{\peval}[3]{{\mathrm{peval}}{\left(#1,#2,#3\right)}}
\newcommand{\pevalp}[3]{{\mathrm{pevalp}}{\left(#1,#2,#3\right)}}
\newcommand{\nullsys}[2]{{\mathrm{nullsys}}{\left(#1,#2\right)}}
\newcommand{\clearf}[1]{{\mathrm{clear}}{\left(#1\right)}}
\newcommand{\flip}[1]{{\mathrm{flip}}{\left(#1\right)}}
\newcommand{\flop}[1]{{\mathrm{flop}}{\left(#1\right)}}
\newcommand{\posform}[1]{{\mathrm{posform}}{\left(#1\right)}}
\newcommand{\fse}[1]{{\mathrm{fse}}{\left(#1\right)}}
\newcommand{\fsefull}[2]{{\mathrm{fse}_{#2}}{\left(#1\right)}}

\newcommand{\george}{\text{peval-safe}}

\newcommand{\addguards}{%
\ \ \ %
\stackrel{\mathclap{\normalfont\mbox{\tiny add
      guards}}}{\Longleftrightarrow}\ \ \ %
}

\title{Semantics of Division for Polynomial Solvers}
\author{Christopher W.~Brown, United States Naval Academy}

\begin{document}
\maketitle

\begin{abstract}  
  How to handle division in systems that compute with logical formulas
  involving what would otherwise be polynomial constraints over the real
  numbers is a surprisingly difficult question.  This paper argues that
  existing approaches from both the computer algebra and computational
  logic communities are unsatisfactory for systems that consider the
  satisfiability of formulas with quantifiers or that perform quantifier
  elimination. To address this, we propose the notion of the
  \emph{fair-satisfiability} of a formula, use it to characterize
  formulas with divisions that are \emph{well-defined},
  meaning that they adequately guard divisions against division by
  zero, and provide a
  \emph{translation algorithm} that converts a formula with divisions
  into a purely
  polynomial formula that is satisfiable if and only if the original
  formula is fair-satisfiable.  This provides a semantics for division
  with some nice properties, which we describe and prove in the paper.
\end{abstract}

\section{Introduction}
\label{section:intro}
The fundamental problem considered in this paper is how
occurrences of division in what would otherwise be first-order
formulas in elementary real algebra should be handled.
Specifically, we consider this question from the perspective of 
tools from
symbolic computing and real algebraic geometry, which assume
\emph{polynomial} constraints --- an assumption violated by divisions
with non-constant denominators.
The easiest approach, of course,
is to simply disallow division.  However, that ship seems to have
sailed: not only do users ask to use division in input formulas, but
common tools like Maple and Mathematica allow division in their inputs
for polynomial solvers, and 
the SMT-LIB allows division in its Non-linear Real Arithmetic (NRA) theory.
Unfortunately, what supporting
division entails, and even what division means, is not straightforward.
Moreover, solutions that make sense for one community may differ from
what is sensible for another community --- a point which is taken up
more fully later in this section.  This
paper examines the problem of supporting division from the perspective
of the computer algebra and computational algebraic geometry
communities, while contrasting that perspective with others, and
proposes a semantics of division that makes sense for
computer algebra and computational algebraic geometry solvers.

The remainder of this section looks at existing approaches to handing
division in solvers for real non-linear arithmetic, and finds each
approach deficient in various ways.  Section~\ref{section:fairsat}
proposes the notion of a formula with divisions being ``fair-SAT'',
which attempts to characterize when formulas are
satisfiable without divide-by-zero games that rely on a particular
definition of what division by zero means.
This leads to the notion of
a ``well-defined'' formula, which is one in which simply clearing
denominators leaves a polynomial formula whose satisfiability is
equivalent to the fair-SATness of the original formula.
Ultimately, we would like a procedure that translates formulas with
divisions to purely polynomial formulas in such a way that
fair-SATness is preserved.  Section~\ref{section:guard} present the
translation procedure, which gives us a practical way to translate
formulas with divisions into purely polynomial formulas that has nice
properties, some of which are explored in Section~\ref{section:property}.
Interestingly, \emph{nullification} of polynomials, which is an
important issue in Cylindrical Algebraic Decomposition based
algorithms for computing with real polynomial constraints, plays an
important role in the translation algorithm.

\subsection{The semantics of division --- existing choices}
The semantics of real algebra, more precisely 
the first-order theory of real closed fields,
are
clearly and unambiguously defined\footnote{In brief, we have the
axioms of an ordered field, augmented with axioms describing that each
non-negative number has a square root, and each odd-degree polynomial
has at least one root.} (see for example
Section 2.2.2 of \cite{passmore:2011}).  However, extending this theory to
allow division is problematic since in first-order logic all
functions are total while, in usual mathematical practice, divisors of
zero are considered to be outside of the division function's domain.
This leaves anyone intending to extend  
real algebra as a first order theory to include division with two
basic options: 
\begin{itemize}
  \item fix an interpretation for $\text{div}(\cdot,0)$, or
  \item leave $\text{div}(\cdot,0)$ uninterpreted.
\end{itemize}
Computational logic systems like Isabelle \cite{Nipkow2002APA}
(see also \cite{Harrison:1998}) and
ACL2 (see \texttt{Unary-/} documentation in \cite{ACL2v8.5}) have chosen the
former solution, and define $\text{div}(x,0) = 0$.\footnote{
One could chose to define $\text{div}(x,0)$ to be something arbitrary
like 42
or $x - 5$, but there are good reasons to choose zero instead.  
\cite{Bergstra2006,Bergstra2007} address this question from the
perspective of formal algebraic specifications of abstract data types,
and they include discussions of various axiomatizations of rationals
and rational complex numbers and how $0^{-1} = 0$ falls out as a
consequence of ``natural'' additions to the commutative-ring-with-$1$ axioms
that describe $(\cdot)^{-1}$, like
$(x\cdot y)^{-1} = x^{-1}\cdot y^{-1}$ and $(x^{-1})^{-1} = x$.
}
SMT-LIB \cite{BarST-SMT-10,BarFT-SMTLIB-10} has chosen the second
approach (see Section~\ref{section:uninterptroubles} below).  Both of these are 
compromises, and if, as I've often heard said, a good compromise
leaves all parties equally unhappy, these are both good compromises.

The choice to assign division by zero the value zero has the advantage
of being simple, and relatively easy to compute with.
For example, the evaluation of a quantifier-free formula with
divisions at a particular assignment of values for variables
(interpretation of constants from the SMT-LIB perspective) is
straightforward. 
On the other hand, it leaves users
with results that may surprise them, for instance that
$
\frac{x^2 + 1}{x^2} \leq 1
$
has a solution, although that solution does not satisfy $x^2 + 1 \leq
x^2$. 

The choice to keep $\text{div}(\cdot,0)$ uninterpreted has some
advantages that we will describe later, but has the disadvantage of
changing the very meaning of ``solve'' to include not only finding
values for real-valued variables (or ``constants'' in SMT-LIB's
terminology), but also finding an interpretation for
$\text{div}(\cdot,0)$. Moreover, users are faced with surprising
results like the fact that $3/x^2 + 2 = 0$ is satisfiable!
(Any interpretation in which $x=0$ and $\text{div}(3,0) = -2$ is a solution.)

It is worth considering what standard mathematical practice would
produce.  My claim is that in standard mathematical practice an
equality/inequality can never be said to be satisfied at an assignment
of values for which directly evaluating the inequality results in division by
zero.  So, for example, $x/x = 1$ is not considered to be satisfied at
$0$. This is not an acceptable view for logicians, though, since it
leads to strange phenomena in logic.  For example,
$\exists x[ 1/x^2 < 0]$ is clearly not satisfied, so we would expect its
negation to be satisfied.  But $\neg \exists x[1/x^2 < 0]$
should be equivalent to $\forall x[1/x^2 \geq 0]$, at least if you
believe the Trichotemy Law, and this formula is also false because it
fails to be satisfied at $x = 0$. Perhaps more importantly, the
axiom of reflexivity for equality fails to hold for terms with denominators
that can vanish!

The point here is that
$\neg \exists x[1/x^2 < 0] \Leftrightarrow  \forall x[1/x^2 \geq 0]$ is a general property of ordered rings,
not anything specific to the reals, and $\forall x[\text{div}(x,0)=\text{div}(x,0)]$ is a
property of first-order logic with equality.  So adopting the usual
mathematical practice that an (in)equality involving divisions is
never satisfied when one or more denominators vanish has the
consequence of invalidating logical theories that are more
fundamental than the real numbers --- the theory of ordered rings,
for example, or even the theory of equality.  This certainly makes it
clear why computational logic systems cannot adopt that practice!

Finally, we turn to what existing practice is in computer algebra
systems.  Mathematica \cite{Mathematica:24} and Maple \cite{Maple:24}
are two of the most used Computer Algebra Systems.
Mathematica's \texttt{Resolve} command provides quantifier elimination
for real algebra, with support for non-constant denominators.
Its approach to handling constraints with denominators is to add
``guards'', i.e. constraints that ensure the non-vanishing of
denominators, and then to clear denominators.  So, for example,
$f/g = 0$ would be replaced with $g \neq 0 \wedge f = 0$
\cite{Strzebonski:2024}.
Maple also provides quantifier elimination for real algebra with its
\texttt{QuantifierElimination} command.  Like Mathematica, division
with non-constant denominators is supported, with essentially the same
approach as Mathematica\footnote{For more details on the guarding
process, see the documentation for the
\texttt{ConvertRationalConstraintsToTarski} command.}. This practice
of adding guards has its own limitations, which we describe more fully
in Section~\ref{section:capractice}.

\subsection{Why supporting existing semantics from logic is undesireable}
In this section we argue that it is undesireable to try to have 
systems from computer algebra or computational algebraic geometry
that compute with real polynomial (in)equalities support division
in a way that mimics either the semantics of Isabelle/ACL2 or those of
SMT-LIB.
First of all, we have reservations about both choices.  They do not, as
some of the above examples show, match common mathematical
practice. Of course, they cannot, because of the restrictions of the
domains these 
systems work in.  So people working in these areas made the compromises
that work best for them. But these are not necessarily compromises
that make sense for computer algebra or computational real algebraic geometry.

\subsection{Problems with the $x/0=0$ approach}
\begin{figure}
  \label{figure:div0is0}
    \begin{center}
      \includegraphics[width=2in]{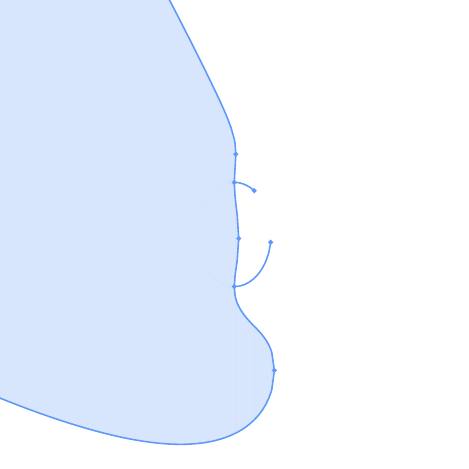}
      \ \ \ \includegraphics[width=2in]{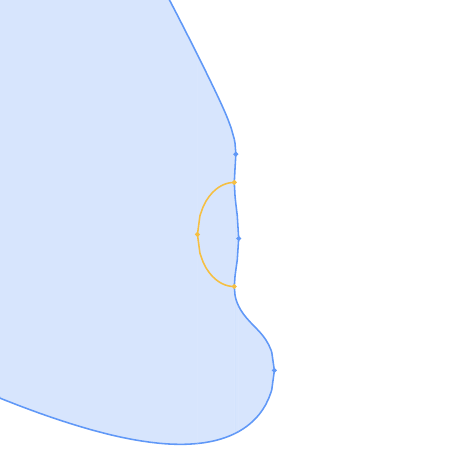}
    \end{center}
    \caption{
      The figure on the left is the set of $(z,y)$-values satisfying
      $F'_1$. The figure on the right is the set of $z,y$-values
      satisfying the $x \neq 0$ branch of formula $F'_1$, which
      corresponds to guarding against division by zero and
      clearing denominators in $F_1$, but without adding the extra
      branch introduced by defining $y/0$ to be zero. The half-ellipse
      cut out in the right figure corresponds to $x = 0$, i.e. where
      the denominator vanishes, which has
      been disallowed.  In the left figure, some but not all of that
      ellipse has been introduce into the solution.  The reason part
      of the ellipse is cut out is that a new constraint,
      $8 y  \leq 8 - (z + y + 1)^2$, is introduced by defining
      $y/0=0$.  This is a constraint (and, underlying it, a
      polynomial) that in usual mathematical practice has no
      connection to the original problem.
    }
\end{figure}

The disadvantage of the divide-by-zero-gives-zero approach,
other than surprising the user with unexpected results as seen above,
is that the
translation process systematically introduces new polynomials that are
artifacts of the $x/0=0$ choice, not of the problem itself, viewed from a natural
mathematical perspective.  If an input formula contains a division
with denominator $D$ in
atom $A\ \sigma\ 0$, then $A\ \sigma\ 0$ must be rewritten as
$D \neq 0 \wedge A_1\ \sigma\ 0 \vee D = 0 \wedge A_2\ \sigma\ 0$,
where $A_1$ is the expanded form of $D^2\cdot A$ and $A_2$ is $A$ with
any term $\text{div}(\cdot,D)$ replaced with 0.  Obviously there are some
situations in which this may be simplified, but in general this is what
will have to be done. The following example, while perhaps not the
simplest, shows how this process introduces new polynomials that would
not otherwise be part of the problem (meaning if standard mathematical
practice were followed). Consider the folowing formula (with one division).
  $$
F_1 := \exists x[ z^2 - 3 x + 2 y^2 = 1 \wedge 8 y + 8 y/x^2 \leq 8 - (z + y + 1)^2
  $$
Translating this into an equivalent formula with only polynomial
constraints gives:
$$F'_1 := \exists x[ z^2 - 3 x + 2 y^2 = 1 \wedge
  [
    x \neq 0 \wedge 8 x^2 y + 8 y \leq  8 x^2 - x^2 (z + y + 1)^2
    \vee
    x = 0 \wedge  8 y  \leq  - (z + y + 1)^2
]]$$
Performing quantifier elimination on $F'_1$ we get the set depicted on
the left in Figure~\ref{figure:div0is0}.  This figure illustrates some of the, perhaps,
surprising results of the $x/0=0$ choice of semantics --- results
which have negative impacts on computation too.  First of all, a
lower-dimensional component to the solution is introduced, which is
not part of what standard mathematical practice would call the
solution.  Second of all, the polynomial $8 y - 8 + (z + y + 1)^2$,
which is only connected to the problem because of the choice to define
$x/0=0$,
is introduced as part of the solution.  If this were part of a larger
problem, we would carry the computatoinal burden of these (from the standard
mathematical perspective) spurrious components and polynomials.
%
Thus, problems that are already computationally challenging will become
still harder, and I suspect that will often be unintentional on the
part of users.

\subsection{Problems with leaving div$(\cdot,0)$ uninterpreted}
\label{section:uninterptroubles}
As noted above, the decision to leave $\text{div}(\cdot,0)$
uninterpreted leads to some results that may be surprising to users.  
However, adopting the SMT-LIB approach to division is actually much
more problematic for solvers from 
computer algebra or computational real algebraic geometry.  The file
\verb|Reals.smt2|, which 
defines the theory of the reals in SMT-LIB, defines division ``as a
total function that coincides with the real division function  
for all inputs x and y where y is non-zero'', along with the
amplifying note:

\begin{verbatim}
  Since in SMT-LIB logic all function symbols are interpreted as total
  functions, terms of the form (/ t 0) *are* meaningful in every 
  instance of Reals. However, the declaration imposes no constraints
  on their value. This means in particular that 
  - for every instance theory T and
  - for every value v (as defined in the :values attribute) and 
    closed term t of sort Real,
  there is a model of T that satisfies (= v (/ t 0)).
\end{verbatim}

This means that an interpretation can have different results for
$\text{div}(x,0)$ for different values of $x$.  In fact, it is
completely free so that 
even statements like $y\cdot\text{div}(x,0) = \text{div}(x\cdot y,0)$
need not hold.  The problem is that this provides division with too
much expressivity for computer algebra and computational geometry
algorithms to support.  Consider the following
examples:

\begin{itemize}
\item Let $F(x_1,x_2,\ldots,x_n)$ be a quantifier-free formula without
  division (i.e. purely polynomial). $F$ 
  is equi-satisfiable with the formula
  $F(\text{div}(1,0),\text{div}(2,0),\ldots,\text{div}(n,0))$.
  Indeed these formulas are equivalent in the sense that satisfying
  interpretations for the two are isomorphic under
  $x_1 = \text{div}(1,0),x_2 = \text{div}(2,0),\ldots,x_n = \text{div}(n,0)$.
  For example $[x^2+ y^2 + 3 x y + (x - y)^4 \leq -1/100 ]$
  is satisfiabile if and only if there is an interpretation for
  $\text{div}(\cdot,0)$ that is a model for
  $$[\text{div}(1,0)^2+ \text{div}(2,0)^2 + 3\cdot \text{div}(1,0)\cdot \text{div}(2,0) + (\text{div}(1,0) - \text{div}(2,0))^4 \leq -1/100 ].$$
  So occurences of div can produce what are, from the perspective of
  symbolic computing and real algebraic geometry, new variables, and of
  course the number of variables is a critical complexity parameter for
  algorithms in this area.  In particular, in the above example
  determining the satisfiability of a formula without variables
  (SMT-LIB constants) requires determining the satisfiability of an
  $n$-variable Tarski formula.

  It's also worth noting that it isn't necessarily easy to identify
  divisions that act as variables in this way.  For example, suppose a
  formula has an (in)equality containing variables $x$ and $y$ and
  terms $\text{div}(x,p(x,y))$, $\text{div}(y,p(x,y))$ and
  $\text{div}(x+y,q(x,y))$, where $p$ and $q$ are polynomials.
  In other words, assume that the (in)equality is of the form
  $$A(x,y,\text{div}(x,p(x,y)),\text{div}(y,p(x,y)),\text{div}(x+y,q(x,y)))
  \ \sigma\ 0.$$
  If $(x,y) = (\alpha,\beta)$ is a point at which 
  $p$ and $q$ have a common zero and at which $x$, $y$ and
  $x+y$ all have distinct values, deciding whether the constraint is
  satisfied at $(\alpha,\beta)$
  requires considering the polynomial
  $A(\alpha,\beta,d_1,d_2,d_3)$, where $d_1,d_2,d_3$ are new
  real-valued variables representing the interpretations
  of $\text{div}(\alpha,0)$, $\text{div}(\beta,0)$ and
  $\text{div}(\alpha+\beta,0)$. The point is, even recognizing that
  there is a three-variable problem lurking in the original formula
  means solving the systems $p(x,y) = q(x,y) = 0$ and checking for
  solutions at which $x$, $y$ and $x+y$ are distinct.  Moreover, the
  three-variable problem we get will, in general, be over the
  algebraic extension required to represent $\alpha$ and $\beta$.

\item Many computer algebra systems and specialized programs from the
  computer algebra community are able to do satisfiability checking
  for formulas with quantifiers or even to do quantifier elimination.
  Consider the problem of deciding the satisfiability of formulas
  that include division following the SMT-LIB semantics
  where quantification is allowed. Let $p$ be an $n$-variable
  polynomial with integer coefficients.  Deciding the satisfiability
  of
  {\small $$
  \forall z [ (0 < z < 1 \Rightarrow \text{div}(z,0) = 0) \wedge
    \text{div}(z+1,0) = \text{div}(z,0) ] \wedge
  \text{div}(x_1,0) = 1 \wedge
  \cdots
  \wedge \text{div}(x_n,0) = 1 \wedge
  p(x_1,\ldots,x_n) = 0
  $$}
  is equivalent to determining whether $p$ has a solution over the
  integers, which is undecidable.

\item In some ways the biggest issue is that some problems that are
  important in computer algebra and computational algebraic geometry
  are not well-defined when we adopt the SMT-LIB semantics.
  For
  example, a common problem is to compute the dimension of a
  semi-algebraic set given a defining formula.  But consider the
  set $S$ defined by
  $S = \{(s,t)\in\realring^2|1/t \leq s \leq 1 \wedge t^2 \leq 0\}$.
  The dimension of $S$ 
  is 1, 0 or $-1$ depending on the interpretation of
  $\text{div}(1,0)$.  Obviously, $(s,t)$ cannot be in $S$ unless
  $t = 0$.  So $S$ is really defined by $\mydiv{1}{0} \leq s \leq 1$.
  Under an interpretation for which $\mydiv{1}{0} < 1$, $S$ is a
  non-degenerate interval on the $t$-axis, and so has dimension one.
  Under an interpretation for which $\mydiv{1}{0} = 1$, $S$ contains
  only the single-point $(1,0)$, and so has dimension zero. Finally,
  under any interpretation for which $\mydiv{1}{0} > 1$, $S$ is empty.
  
  Another common problem with similar issues is the question of
  whether the set of points satisfying a formula is connected.  For
  example, when there are two components connected by a single point,
  the presence or absence of that point in the set defined by a
  formula may depend on the interpretation of $\mydiv{\cdot}{0}$.
\end{itemize}

So, supporting the SMT-LIB semantics for division is not practical for
systems that are intended to compute with polynomials because it means
that the presence of division
can cause problems to jump in complexity, to switch from
decidable to undecidable, or to cease to be well defined.  Even if we
consider the above examples to be ``misuses'' of 
division, it is not at all clear (to me, at least) that one can detect
whether an input formula has divisions that cause these kind of
problems. 

\subsection{Problems with adding guards following existing computer algebra practice}
\label{section:capractice}
The practice of adding guards as described for Mathematica's
\texttt{Resolve} and Maple's \texttt{QuantifierElimination} commands is
pragmatic --- but it doesn't provide a completely satisfactory
semantics.  There are instances of formulas that are equivalent based
on fundamental laws like trichotemy\footnote{The Trichotemy law is an
axiom of ordered rings stating that any ring element is exactly one of
positive, negative or zero.  As a consequence,
$\neg[f < g] \Longleftrightarrow [f \geq g]$ and the other similar
identities hold for any ordered ring, not just the reals.
} or substitutivity of equality,
for which adding guards produces inequivalent formulas.

For example, if we assume the Trichotemy Law holds, the formula
$F := \exists x[ 1/x^2 < 0]$ has the property that the guarded
versions of both $F$ and $\neg F$ are false.
$$
\begin{array}{rl}
  \exists x[ 1/x^2 < 0]&
  \hspace{12pt}
  \stackrel{\mathclap{\normalfont\mbox{\tiny add guards}}}{\Longleftrightarrow}
  \hspace{12pt}  
  \exists x [ x \neq 0 \wedge 1 < 0]
  \hspace{12pt}  
  \Longleftrightarrow
  \hspace{12pt}     
  \false\\
  \neg \exists x[ 1/x^2 < 0]&
  \hspace{12pt}     
  \stackrel{\mathclap{\normalfont\mbox{\tiny trichotemy law}}}{\Longleftrightarrow}
  \hspace{12pt}     
  \forall x [ 1/x^2 \geq 0]
  \hspace{12pt}     
  \stackrel{\mathclap{\normalfont\mbox{\tiny add guards}}}{\Longleftrightarrow}
  \hspace{12pt}     
  \forall x [ x \neq 0 \wedge 1 \geq 0]
  \hspace{12pt}     
  \Longleftrightarrow
  \hspace{12pt}     
  \false
\end{array}
$$
A subtle effect of this kind of issue is that the result of the guarding process
depends on when it happens with respect to other, even very basic, rewriting steps.
For instance, in the second line above, had we done the
guarding prior to moving the negation inwards we would have gotten
$$
\neg \exists x[ 1/x^2 < 0]
\ \ \ \stackrel{\mathclap{\normalfont\mbox{\tiny add
      guards}}}{\Longleftrightarrow}\ \ \ \
\neg \exists x[ x \neq 0 \wedge 1 < 0]
\ \ \ \Longleftrightarrow\ \ \ \forall x [ x = 0 \vee 1 \geq 0]
\ \ \ \Longleftrightarrow\ \ \ \true
$$
... which is the opposite result.

In computer algebra solvers it is common to require prenex inputs, and
this issue of when the guarding happens becomes an issue when we
consider when the prenex conversion happens:
$$
\begin{array}{rcl}
\forall y[\neg\forall x[1/x^2 \geq 1/y] \vee y < 0]
& \addguards &
\forall y[\neg\forall x[x \neq 0 \wedge y \neq 0 \wedge 1/x^2 \geq 1/y] \vee y < 0]\\
&&\forall y[\exists x[x = 0 \vee y \neq 0 \vee 1/x^2 < 1/y] \vee y < 0]\\
&&\forall y \exists x [x = 0 \vee y = 0 \vee y^2 < x^2 y \vee y < 0]\\
&&\true\\
\forall y[\neg\forall x[1/x^2 \geq 1/y] \vee y < 0]
&\Longleftrightarrow &\forall y[\exists x[1/x^2 < 1/y] \vee y < 0]\\
& \addguards &
\forall y[\exists x[x \neq 0 \wedge y \neq 0 \wedge y^2 < x^2 y] \vee y < 0]\\
&&\forall y\exists x[x \neq 0 \wedge y \neq 0 \wedge y^2 < x^2 y \vee y < 0]\\
&&\false
\end{array}
$$
So we get different results if the guarding is done before versus
after the conversion to prenex.

If we assume reflexivity and 
substitutivity of equality, then
$\forall x, y[ x = y \Rightarrow 1/x = 1/y ]$
is true, but
$$
\forall x, y[ x = y \Rightarrow 1/x = 1/y ]
\ \ \ %
\stackrel{\mathclap{\normalfont\mbox{\tiny add
      guards}}}{\Longleftrightarrow}\ \ \ %
\forall x, y[ x = y \Rightarrow x \neq 0 \wedge y \neq 0 \wedge y = x ]
$$
... is false, since the implication fails at $x = y = 0$.

Finally consider the following example of two formulas that are
equivalent given only that ``$\neq$'' is the negation of ``$=$'', and
yet give different results under the guarding regime:
$$
\exists x[1/x = 0] 
\ \ \ %
\stackrel{\mathclap{\normalfont\mbox{\tiny add
      guards}}}{\Longleftrightarrow}\ \ \ %
\exists x[x \neq 0 \wedge 1 = 0]
\Longleftrightarrow
\false
$$
$$
\exists x[\neg (1/x \neq 0)] 
\ \ \ %
\stackrel{\mathclap{\normalfont\mbox{\tiny add
      guards}}}{\Longleftrightarrow}\ \ \ %
\exists x[\neg(x \neq 0 \wedge 1 \neq 0)]
\Longleftrightarrow
\exists x[x = 0 \vee 1=0]
\Longleftrightarrow
\true
$$
In this example, $\exists x[1/x = 0]$ leads to $\false$, which seems
clearly correct.  One might be tempted to suggest that the guarding
works as long as any logical negations are pushed into the
inequalities before doing the guarding.  But one can also construct
examples where this does not work. Consider:
$$
\exists x \forall y[ y^2 > 1 \vee \neg ( (x/y)^2 < 4 ) ] 
\addguards
\exists x \forall y[ y^2 > 1 \vee \neg ( y \neq 0 \wedge x^2 < 4 y^2) ] 
\Longleftrightarrow
\true \text{ (e.g. at $x=2$)}
$$
$$
\exists x \forall y[ y^2 > 1 \vee (x/y)^2 \geq 4 ] 
\addguards
\exists x \forall y[ y^2 > 1 \vee ( y \neq 0 \wedge x^2 \geq 4 y^2) ] 
\Longleftrightarrow
\false
$$
I would argue that the former has a far better claim to being correct
than the latter, so pushing the logical negations into the
inequalities first did not work for this example, although it did for
the previous example.

The issues just described are serious.  They show that this simple
guarding procedure conflicts with
the theory of ordered rings, and with the rules of first-order
logic with equality, and even with propositional logic plus the
definition of $\neq$ as the negation of $=$.
In fact, I would argue that this is not really a well-defined
semantics --- there is no clear connection between a formula with
divisions and the purely polynomial formula produced from it via the
guarding procedure other than the procedure itself.

\section{What \emph{can} be done about division?}
\label{section:fairsat}
The previous section argued that supporting division with the
semantics of ``usual mathematical practice'', or the $x/0=0$
semantics, or the unterpreted function semantics for
$\text{div}(x,0)$,
or the usual computer algebra semantics
is not reasonable.  The question then is what is reasonable?  What
should we be looking for?

I am proposing that we should be looking for a procedure that
translates a formula with divisions into one or more formulas without
divisions.  I propose treating division by zero as uninterpreted, but
I do not propose following SMT-LIB semantics (for the reasons
described above).  Instead, I propose that for \emph{well-defined}
formulas (what that means will be explored later), we guarantee
\emph{equivalence} (once again, a precise definition of what that means comes
later) between the input formula and the result of the
translation process, while for formulas that are not well defined,
we do not provide this guarantee.  However, for formulas that are not
well defined, we define the notion of \emph{fair-SAT}, and we guarantee
that the original formula with divisions is fair-SAT if and only if
the translated formula is SAT.  The idea behind fair-SAT is that a
formula is fair-SAT if and only if it is satisfiable without having 
to rely on divide-by-zero shenanigans ...
a more formal definition appears in Section~\ref{section:welldef2}.

Finally, the mathematical practice of providing a ``guard'' against
division by zero, like the ``$x\neq 0$'' in
$\exists x[x \neq 0 \wedge x + 1/x = 0]$,
should produce formulas that meet the definition of
\emph{well-defined}, as we assume most people would expect.
Moreover, when an inequality is appropriately guarded,
clearing denominators should not change the meaning of the formula.

\subsection{``Well-defined formula'' ... first attempt}
\label{section:firstattempt}
We would like a good notion of formulas containing divisions being
``well-defined''.  This is important because we might hope that for
well-defined formulas we can precisely describe what
we are trying to accomplish --- for example we would like to say that
if $F$ is a well-defined formula, then our translation to a formula
without divisions should be equivalent to $F$.
In first-order logic, all functions are total, so we cannot express
that division by zero is ``undefined''.  We can, however, express that
the satisfiability of a formula is not dependent on the choice of
what value to assign to the result of dividing by zero.  That
motivates our first attempt at pinning down the concept of a
``well-defined'' formula.

\begin{definition}
  \label{definition:fisttry}
A formula $F$ in the theory of real algebra with uninterpreted
function symbol $\mydiv{\cdot}{\cdot}$ and additional axiom
$$
\forall x,y[ y \neq 0 \Rightarrow \mydiv{x}{y} \cdot y = x]
$$
is called \emph{well-defined} if and only if
the true/false value of $F$ is the same under 
under any two interpretations that differ only in the interpretation 
of $\mydiv{\cdot}{0}$.
Note: Other than $\mydiv{\cdot}{0}$, the only thing left to
interpret are the values of the free variables.
\end{definition}

Under this definition, for example, the formula
$\exists x,y[ x-y \neq 0 \wedge x + y =  3 x/(x - y) \wedge y + 1  = 2 x]$
is well-defined (it is false).
On the other hand, $\exists x,y[ x + y =  3 x/(x - y) \wedge y + 1  = 2x]$
is not well-defined according to this definition, since under the interpretation
in which  $\text{div}(3,0) = 2$, the formula is true (satisfied at
$(x,y) = (1,1)$), but for all other interpretations it is false.  This
definition of ``well-defined'' has some nice properties.  For example,
it is immediate that under transformations from first-order logic and
ordered fields that produce equivalent formulas, well-definedness is
maintained. Moreover, it is clear that formulas written with
``guards'', in the usual sense, are well defined.  For example:
$\exists x,y[ y \neq 0 \wedge x^2/y^2 + 2 x/y + 1  = 0]$ is 
well-defined.

Unfortunately, this definition of well-defined causes some problems.
Specifically, we will look at two simple formulas --- both closed with
only an existential quantifier block, and both well defined under
Defintition~\ref{definition:fisttry} --- and show that whatever
combination of guarding and denominator-clearing approaches we take,
the resulting formula is not equivalent to the original under any
interpretation for $\mydiv{\cdot}{0}$.

Let $F_1$ and $F_2$ denote the following formulas:
$$
F_1 := \exists x,y\left[y^2\left(1 + x^2 + \frac{1}{y^2}\right) \leq 0)\right],
F_2 := \exists x\left[ 0 \leq x \wedge \frac{1}{x} \leq \frac{1}{x}- 1\right]
$$
and note that
\begin{itemize}
\item $F_1$ and $F_2$ are well-defined by Definition~\ref{definition:fisttry},
  and
\item $F_1$ is true under any interpretation for
  $\text{div}(\cdot,0)$, and $F_2$ is false under any interpretation
  for   $\text{div}(\cdot,0)$.
\end{itemize}

For fully existentially quantified conjunctions, there are two basic
approaches --- clear denominators but add guards to guarantee that the
conjunction is false when a denominator is zero, or clear denominators
without adding guards.  Additionally, sometimes denominators can be
cleared simply by ``expanding'', whereas sometimes one has to multiply
both sides of an inequality by the square of the denominator.  We
consider these possiblities below:

\begin{enumerate}
\item Expanding to clear denominators, without adding ``guards'':
  for $F_1$ this produces formula\\
  $\exists x,y\left[y^2 + y^2 x^2 + 1 \leq 0\right]$
  which is false.
\item Expanding to clear denominators, and adding ``guards'':
  for $F_1$ this produces 
  formula\\
  $\exists x,y\left[y \neq 0 \wedge y^2 + y^2 x^2 + 1 \leq 0\right]$
  which is false.
\item Multiplying both sides by the denominator squared and
  expanding, without adding ``guards'':
  for $F_2$ this produces
  $\exists x\left[ 0 \leq x \wedge x \leq x- x^2\right]$, which is true.
\item Multiplying both sides by the denominator squared and
  expanding, and adding ``guards'': for $F_1$ this 
  produces formula
  $\exists x,y\left[y \neq 0 \wedge y^4(y^2 + y^2 x^2 + 1) \leq 0\right]$
  which is false.
\end{enumerate}

So the moral of the story, is that, under
Definition~\ref{definition:fisttry},  no matter which of the obvious rewriting strategies
you employ, there will be examples of well-defined formulas for which
the rewriting produces a formula that is not equivalent to the
original formula.

It is also interesting to note that recognizing that a formula is
well-defined according to this definition can be quite difficult.
In fact, through a variant of the second example from
Section~\ref{section:uninterptroubles}, it can shown to be undecidable.

\subsection{``Well-defined formula'' ... second attempt}
\label{section:welldef2}
Our first attempt at a definition for ``well-defined'' was based on
the idea of a formula's evaluation at an interpretation not depending
on what \emph{real values} the interpretation assigns to \emph{divisions} with
zero denominators.  
In our second attempt at a servicable definition of a well-defined
formula, we focus on the idea of a formula's evaluation at an
interpretation not depending 
on what \emph{boolean values} are assigned to \emph{atomic formulas}
when they include
divisions with zero denominators.
This will lead us to the definition of well-defined we will stick
with, and later we will provide a procedure for translating formulas
with divisions into Tarski formulas that maintains equivalence for
well-defined formulas.

\subsubsection{Partial evaluation: peval}
First we require a careful description of the (partial)
evaluation of an atomic formula when divisions appear.  The reason
some care is required is that we must be sure to detect division by
zero.  So, for example, when computing the partial evaluation of
$x\cdot\mydiv{1}{y} = 0$ at $x=0$, the result must be
$0\cdot\mydiv{1}{y} = 0$, not $0=0$.  The reason is that if the
evaluation continues with $y=0$, we must detect that the denominator
has vanished.

\begin{definition}[partial evaluation of a term]
$\peval{T}{x}{\alpha}$,
the partial evaluation of term $T$ at $x = \alpha$, where $x$ is either
a variable or a non-literal constant and $\alpha$ is a real number
is (note: we introduce $\fail$ as a new pseudo-term):
\begin{enumerate}
\item if $T$ does not contain term $x$, $\peval{T}{x}{\alpha}$ is $T$
\item if $T = X$, $\peval{T}{x}{\alpha}$ is $\alpha$.
\item if $T = -T_1$, then
  let $S_1 = \peval{T_1}{x}{\alpha}$.
  If $S_1$ is $\fail$, the result is $\fail$.
  If $S_1$ is a real constant, the result is the real
  constant resulting from computing $-S_1$.
  Otherwise, the result is the term $-S_1$.
\item if $T = T_1\ op\ T_2$, where $op$ is +,- or $*$, then
  let $S_1 = \peval{T_1}{x}{\alpha}$ and $S_2 = \peval{T_2}{x}{\alpha}$.
  If either $S_1$ or $S_2$ is $\fail$, the result is $\fail$.
  If both $S_1$ and $S_2$ are real constants, the result is the real
  constant resulting from computing $S_1\ op\ S_2$.
  Otherwise, the result is the term $S_1\ op\ S_2$.
\item if $T = \mydiv{T_1}{T_2}$, then
  let $S_1 = \peval{T_1}{x}{\alpha}$ and $S_2 = \peval{T_2}{x}{\alpha}$.
  If either $S_1$ or $S_2$ is $\fail$, the result is $\fail$.
  If $S_2$ is the real constant $0$, the result is $\fail$.
  If both $S_1$ and $S_2$ are real constants, $S_2\neq 0$, the result
  is the real constant resulting from computing $S_1 / S_2$.
  Otherwise, the result is the term $\mydiv{S_1}{S_2}$.
\end{enumerate}
\end{definition}

A few examples of the evaluation of different terms will, hopefully,
clarify what is being done here:
$$
\begin{array}{rrcl}
a.& \peval{x\cdot\mydiv{1}{y}}{x}{0} &\longrightarrow& 0\cdot\mydiv{1}{y}\\
b.& \peval{x\cdot\mydiv{1}{y}}{y}{0} &\longrightarrow& \fail\\
c.& \peval{(x\cdot 2) - y}{x}{3/4} &\longrightarrow& 3/2-y\\
\end{array}
$$

Moving forward, it will be convenient to be able to express the
p-evaluation of a term (and later a formula) at a sequence of variable-value pairs as a
single operation rather than an awkward nesting of p-evaluations at
individual variable-value pairs.

\begin{definition}[p-evaluation extended to tuples]
  Let $(x_{i_1}, x_{i_2}, \ldots, x_{i_t})$ be a
  $t$-tuple of variables, and 
  $\gamma \in \realring^t$, and $w$ be a term (or, later, a formula).
  Define
  $$
  \pevalp{w}{(x_{i_1},\ldots,x_{i_t})}{\gamma} = \left\{
  \begin{array}{cl}
    w & \text{if $t < 1$,}\\
    \peval{ \pevalp{w}{(x_{i_1},\ldots,x_{i_{t-1}})}{(\gamma_1,\ldots,\gamma_{t-1})} }{x_{i_t}}{\gamma_{t}}
    & \text{otherwise.}
  \end{array}
  \right.
  $$
\end{definition}

It will be useful in what follows to have a way to refer to
evaluations that do or do not result in vanishing denominators.  This
is complicated by the fact that peval is a ``lazy'' procedure,
meaning that it delays arithmetic operations one might normally move
forward with early.  For example, normally when we see $x\cdot 0$, we
immediately evaluate to $0$.  However, pevalp delays this step until
$x$ has a value.  We call a partial evaluation ``legal'' if there is
no division by zero, \emph{and} there is a way to continue evaluating
the remaining variables that does not produce a division by zero.  So
a partial evaluation is ``illegal'' only if it involves a division by
zero, or will inevitably lead to a division by zero.  This is made
precise in the following defintion.

\begin{definition}
  \label{definition:legal}
  Let $T$ be a term, possibly with $\mydiv{\cdot}{\cdot}$
  terms, and let $x_{i_1} \prec \cdots \prec x_{i_n}$ be an ordering of
  variables that includes the variables in $T$. Consider
  $\gamma\in\realring^t$, where $t \leq n$.
  [Note, we will use ``:'' to denote concatenation of tuples.]
  We say \emph{$T$ is legal at $\gamma$} if
  there exists $\beta\in\realring^{n-t}$ such that
  $\pevalp{T}{(x_{t_1},\ldots,x_{t_n})}{\gamma:\beta}$ is not $\fail$,
  and \emph{illegal} otherwise.
\end{definition}

We next consider extending the concept of partial evaluation to
formulas.  The main issue in doing this is how to handle an atomic
formula $f\ \sigma\ g$ when partial evaluation of either $f$ or $g$
results in $\fail$.  In this case, we define the evaluation of the
constraint to result in a fresh\footnote{A ``fresh'' variable is
one that has not appeared at any point, i.e. different than any
previously used variable.} propositional variable $U_i$.  Later in the
paper, all $U_i$ variables will be universally quantified, following
the idea stated at the beginning of this section that a
``well-defined'' formula is one that is true regardless of what
true/false value we give to constraints with vanishing denominators.
So, for example, the definition below states that
$$
\peval{\exists r \forall y[ x \geq r \vee \mydiv{y^2}{x-1} \leq 0]}{x}{1}
\longrightarrow
\exists r \forall y[1 \geq r \vee U_1].
$$
Previewing ideas that are defined precisely later:
if we consider $r=2$ we get
$\forall y[\false \vee U_1]$, for which it is not the case that the
formula is true for all values of $U_1$, but if we choose $r=0$
we get
$\forall y[\true \vee U_1]$, for which it is the case that the
formula is true for all values of $U_1$ --- which is sufficient since
$r$ is existentially quantified.
So we consider the original formula to be true at $x=1$ despite the
vanishing denominator.

Later we will introduce ``$V_i$'' propositional variables that are
counter-parts to the $U_i$ variables.  Their roles will be explained
later, but because partial evaluation may encounter either kind of
variable, they accounted for in the definition below, even though
their purpose has not yet been described.

\begin{definition}[partial evaluation of a formula]
Let $H$ be a formula, possibly with divisions, and possibly with
propositional variables $U_i$ and $V_j$, where $i$ and $j$ are natural
numbers, in which names have been standardized apart.
$\peval{H}{x}{\alpha}$, the
\emph{partial evaluation of $H$ at $x = \alpha$}, where $x$ is 
a variable free in $H$ 
and $\alpha$ is a real number is
defined by the following rules:
\begin{enumerate}
\item if $H$ is a boolean constant or propositional variable,
  $\peval{H}{x}{\alpha} = H$.
\item if $H$ is the atom $f\ \sigma\ g$. In the case that one or both
  of $f$ and $g$ are illegal at $x = \alpha$,
  the result is a fresh propositional variable $U_i$ (note that
  different occurrences of the same atom get different propositional
  variables). 
  In the case that $f$ and $g$ are both legal at $x = \alpha$, let
  $f' = \peval{f}{x}{\alpha}$ and $g'=\peval{g}{x}{\alpha}$.
  If both $f'$ and $g'$ are real constants, the 
  result is the boolean value of the real number comparison $f'\ \sigma\ g'$.
  Otherwise the result is the predicate expression $f'\ \sigma\ g'$.
\item if $H$ is $H_1\ op\ H_2$, where $op$ is a binary boolean operator,
  let $H'_1 = \peval{H_1}{x}{\alpha}$ and
      $H'_2 = \peval{H_2}{x}{\alpha}$,
  the result is the formula $H'_1\ op\ H'_2$.
\item if $H$ is $\neg H_1$, 
  let $H'_1 = \peval{H_1}{x}{\alpha}$.
  the result is the formula $\neg H'_1$.
\item if $H$ is $\forall y[H_1]$ (respectively $\exists y[H_1]$),
  then if $y$ is $x$, the result is $H$. Otherwise,
  let $H'_1 = \peval{H_1}{x}{\alpha}$,
  the result is the formula $\forall y[H'_1]$ (respectively $\exists y[H'_1]$).
\end{enumerate}
\end{definition}

A few examples of the evaluation of different formulas will, hopefully,
clarify what is being done here:
$$
\begin{array}{rrcl}
  a.& \peval{%
  x\cdot\mydiv{1}{y} > 0 \wedge x < 5 \vee U_1}{x}{0}
  &\longrightarrow& 0\cdot\mydiv{1}{y} > 0 \wedge \true{} \vee U_1\\
  b.& \peval{%
  x\cdot\mydiv{1}{y} > 0 \wedge x < 5 \vee U_1}{y}{0}
  &\longrightarrow&U_2 \wedge x < 5 \vee U_1\\
  c.& \peval{%
  \mydiv{1}{y} > 0 \Rightarrow \mydiv{1}{y} > 0}{y}{0}
  &\longrightarrow&U_1 \Rightarrow U_2\\
  c.&
  \peval{%
    \peval{x\cdot (y^2-1)\neq 0 \wedge y\cdot \mydiv{1}{x} < 3}{y}{0}}{x}{0}

  &\longrightarrow& \false{} \wedge U_1   \\
  d.&
  \peval{\exists x[ \mydiv{x}{y} > 1 \wedge \mydiv{x}{y-3} > x^2]}{y}{3}
  &\longrightarrow& \exists x[ \mydiv{x}{3} > 1 \wedge U_1]   \\
\end{array}
$$

\subsubsection{Fair-SAT}
In what follows, we will be starting with formulas in which the
variables are standardized apart and $\wedge/\vee/\neg$ are the only
boolean operators and repeatedly peval-uating.  
It is convenient to have a name for formulas of the form that arise
from this process.

\begin{definition}
  We say closed formula $F$ is \emph{\george{}} if the real variables
  are standardized 
  apart, the only boolean operators are $\wedge$, $\vee$ and $\neg$, 
  and for some non-negative integer constant $\#F$,
  there are distinct integers $k_1,\ldots,k_{\#F}$
  such that formula $F$ has
  exactly one occurrence of propositional variable $U_{k_i}$ or $V_{k_i}$, but
  not both, and these are the only propositional variables.  Note that
  divisions and true/false constants are allowed.
\end{definition}

So, for example,
$\exists x[ U_1 \wedge x^2 < 2] \vee \exists y[\mydiv{2,y} = 0 \wedge U_2]$
is \george{}, while the very similar formula
$\exists x[ U_1 \wedge x^2 < 2] \vee \exists x[\mydiv{2,x} = 0 \wedge U_2]$
is not, since the variables are not standardized apart. Also, the
$\exists x[ U_1 \wedge x^2 < 2] \vee \exists y[\mydiv{2,y} = 0 \wedge V_1]$
fails to be \george{} because it has two variables with the
same subscript.

A common operation in what follows will be ``flipping'' the $U_i$s to
$V_i$s, or combining that with logical negation, which we call
``flopping''.
This arises when we consider negations of \george{} formulas.  For
example, we consider the formula $F := \forall x[ x^2 \geq 0 \wedge U_1]$
to be ``false'' (more precisely, it is not ``fair-SAT'', a concept defined
shortly), because it is not true for both
values of $U_1$.  However, the usual negation of $F$ gives
$\exists x[ x^2 < 0 \vee \neg U_1]$, which is also not satisfied
for both values of $U_1$.  
So instead we both negate logically \emph{and} 
flip $U_1$ to $V_1$ --- a combination we call ``flop''.  This
gives
$\exists x[ x^2 < 0 \vee \neg V_1]$, which we do consider to be ``true''
(more precisely, ``fair-SAT''),
since the $V_1$ is interpreted as existentially not universally quantified.
We next provide precise definitions of ``flop'' and ``flip'', which
are then used in the definition of fair-SAT.
\vspace{9pt}
\\
\textbf{Note:} for the remainder of this paper, we will adopt the
convention that $\overline{\sigma}$, where $\sigma$ is a relational
operator, is that operator's negation, i.e.
if $\sigma$ is $<$ then $\overline{\sigma}$ is $\geq$, 
if $\sigma$ is $\neq$ then $\overline{\sigma}$ is $=$
and so on.  Similarly, if $\overline{X}$, where $X$ is either $\exists$
of $\forall$, denotes the opposite of whichever quantifier $X$ is.

\begin{definition}
  Let formula $F$ be \george{}. $\flip{F}$ and $\flop{F}$ are defined
  as follows:
  \begin{itemize}[noitemsep,topsep=3pt,parsep=3pt,partopsep=0pt]
    \item $\flop{\true} = \false$,
      $\flop{\false} = \true$,
      $\flip{\true} = \true$,
      $\flip{\false} = \false$.
    \item
      $\flop{U_i} = \neg V_i$,
      $\flop{V_i} = \neg U_i$,
      $\flip{U_i} = V_i$,
      $\flip{V_i} = U_i$.
    \item
      $\flop{f\ \sigma\ g} = f\ \overline{\sigma}\ g$, 
      $\flip{f\ \sigma\ g} = f\ \sigma\ g$.
    \item 
      $\flop{\neg F} = \flip{F}$, $\flip{\neg F} = \flop{F}$.
    \item 
      $\flop{G\wedge H} = \flop{G} \vee \flop{H}$,
      $\flip{G\wedge H} = \flip{G} \wedge \flip{H}$.
    \item 
      $\flop{G\vee H} = \flop{G} \wedge \flop{H}$,
      $\flip{G\vee H} = \flip{G} \vee \flip{H}$.
    \item
      $\flop{\forall x[F]} = \exists x[\flop{F}]$,
      $\flip{\forall x[F]} = \forall x[\flip{F}]$.
    \item
      $\flop{\exists x[F]} = \forall x[\flop{F}]$,
      $\flip{\exists x[F]} = \exists x[\flip{F}]$.
  \end{itemize}
\end{definition}

Now we are in a position to define the 
\emph{fair-satisfiability} of a formula, which makes precise the idea
of the truth or falsity of a formula not depending on
divide-by-zero shennanigans.

\begin{definition}[fair-SAT]
  The following rules define the formulas that are \emph{fair-SAT}.
  Note that the definition is restricted to closed formulas (i.e. no
  free variables).
  \begin{enumerate}
  \item A formula containing only propositional variables,
    true/false constants and boolean operators is \emph{fair-SAT}
    if and only if
    $\forall U_{i_1},\ldots,U_{i_a}
    \exists V_{j_1},\ldots,V_{j_b}[F]$, where
    $U_{i_1},\ldots,U_{i_a}$ are the $U$-variables and
    $V_{j_1},\ldots,V_{j_b}$ the $V$-variables in $F$.
    Note: the following rules only apply if this rule does not apply!
  \item $\neg F$, where $F$ contains quantifiers, is \emph{fair-SAT}
    if and only if
    \begin{enumerate}
      \item $\neg F = \neg (A \wedge B)$ and $\neg A \vee \neg B$ is \emph{fair-SAT}
      \item $\neg F = \neg (A \vee B)$ and $\neg A \wedge \neg B$ is \emph{fair-SAT}
      \item $\neg F = \neg \neg F'$ and $F'$ is \emph{fair-SAT}
      \item
        $\neg F = \neg \exists x[F']$ and $\forall x[\neg F']$ is fair-SAT
      \item
        $\neg F = \neg \forall x[F']$ and $\exists x[\neg F']$ is fair-SAT
    \end{enumerate}    
  \item $F_1 \vee F_2$ is \emph{fair-SAT} if and only if at least one of
    $F_1$, $F_2$ is \emph{fair-SAT}.
  \item $F_1 \wedge F_2$ is \emph{fair-SAT} if and only if both
    $F_1$ and $F_2$ are \emph{fair-SAT}.
  \item Formula $\exists x[F]$ is \emph{fair-SAT} if and only if there
    is a value $\alpha$ for $x$ such that 
    $\peval{F}{x}{\alpha}$ is \emph{fair-SAT}.
  \item Formula $\forall x[F]$ is \emph{fair-SAT} if and only if
    $\exists x[\flop{F}]$ is not \emph{fair-SAT}
  \item Nothing else is \emph{fair-SAT}.
  \end{enumerate}
\end{definition}

\begin{definition}[fair-SAT equivalence]
  Two closed formulas are called \emph{fair-SAT equivalent}
  provided either both are fair-SAT or neither are fair-SAT.
  Two formulas $F_1$ and $F_2$ with free variables $x_1,\ldots,x_k$
  are \emph{fair-SAT equivalent} provided that for all
  $\alpha\in\realring^k$,
  $\peval{F_1}{(x_1,\ldots,x_k)}{\alpha}$ and 
  $\peval{F_2}{(x_1,\ldots,x_k)}{\alpha}$ are fair-SAT equivalent.
\end{definition}
  
\begin{example}
  Let $F$ be 
  $\exists x,y[ x + y =  \mydiv{3 x}{x - y} \wedge y + 1  = 2x]$.
  By point 5 of the definition, for $F$ to be fair-SAT 
  there would have to be some value $\alpha$ we could give to $x$ for
  which
  $$F' = \peval{F}{x}{\alpha} = \exists y[ \alpha + y =  \mydiv{3 \alpha}{\alpha - y} \wedge y + 2\alpha]$$
  is fair-SAT.  This in turn would require 
  some value $\beta$ we could give to $y$, which presents two cases:
  \begin{enumerate}
  \item $\beta = \alpha$: in this case
    $F'' = \peval{F'}{y}{\alpha} = U_1 \wedge \alpha + 1 = 2\alpha$, which
    is not fair-SAT, regardless of the value of $\alpha$, because
    $F''$ is false when $U_1 = \false$.
  \item $\beta \neq \alpha$:
    in this case
    $F''' = \peval{F'}{y}{\beta} =  \alpha + \beta =  \mydiv{3 \alpha}{\alpha - \beta} \wedge \beta + 1 = 2\alpha$
    which, since $\beta \neq \alpha$ is
    $(\alpha + \beta)(\alpha - \beta) =  3 \alpha \wedge \beta + 1 = 2\alpha$,
    and one can verify this is false for any real values of
    $\alpha$ and $\beta$.
  \end{enumerate}
  Thus, $F$ is not fair-SAT.  It is interesting to note that this
  formula is satisfiable for the 
  interpretation in which $\mydiv{3}{0} = 2$ (for $x = y = 1$)).  So
  for SMT-LIB, $F$ is SAT.  
\end{example}

\begin{example}
  Let $F$ be
  $\forall y [ y^2 (1 + 1/y^2) > 0]$.
  By point 6 of the definition, $F$ is fair-SAT if and only if
  $\exists y [ y^2 (1 + 1/y^2) \leq 0 ]$ is not fair-SAT.
  So, for $F$ to fail to be fair-SAT 
  there would have to be some value $\beta$ we could give to $y$ for
  which $\beta^2 (1 + 1/\beta^2) \leq 0$ is fair-SAT,
  which presents two cases:
  \begin{enumerate}
  \item $\beta = 0$: in this case
    $F' = \peval{y^2 (1 + 1/y^2) \leq 0}{y}{\beta} = U_1$, which
    is not fair-SAT because
    $F'$ is false when $U_1 = \false$.
  \item $\beta \neq 0$:
    in this case
    $F'' = \peval{y^2 (1 + 1/y^2) \leq 0}{y}{\beta} = \beta^2(1 + 1/\beta^2) \leq 0$
    which, is false for any non-zero $\beta$.
  \end{enumerate}
  Thus, $F$ is fair-SAT. It is interesting to note that this formula 
  is UNSAT under any interpretation of $\mydiv{1}{0}$.
\end{example}

\subsubsection{A brief discussion of properties of fair-SAT}
\label{section:fairSatProperties}
Hopefully the definition of fair-SAT seems defensible, at least given
the motivation of wanting the truth or falsity of formulas not to
depend on vanishing denominators.  In this section we highlight some
attractive features of the definition of fair-SAT we have chosen.
Proofs of these properties, which are a bit involved, are presented
later.  Specifically, fair-SATness is preserved by the kinds of basic
first-order logic rewritings that we saw in Section~\ref{section:intro} can
cause problems for the usual computer algebra system practice of
guarding divisions with conditions on the non-vanishing of denominators.

\begin{enumerate}[noitemsep,topsep=3pt,parsep=3pt,partopsep=0pt]
\item If $F$ is fair-SAT, then $\neg F$ is not fair-SAT. 
\item Fair-SATness is preserved when rewriting
  inequality $f\ \sigma\ g$ as $\neg[f\ \overline{\sigma}\ g]$, and vice
  versa. 
\item Fair-SATness is preserved across applications of De Morgan's laws.
\item Fair-SATness is preserved when moving
  negations through quantifiers.  I.e. $\neg\exists x[F]$ is fair-SAT
  equivalent to $\forall x[\neg F]$.
\end{enumerate}

Note that by putting these together we see that the usual process for
converting a formula to prenex form preserves fair-SATness.
So, for example, if we are given the formula
$$
\forall x[ \neg[0 \leq 1/x + 2 \wedge 1/x - 2 \leq 0]
    \vee \exists y[ y^2 < 5 \wedge 3 x^2 - 1/y
  + 1 < 0 ]]
$$
and we convert to prenex, simplifying away the negations as we go, we get
$$
\forall x[ \exists y[ 0 > 1/x + 2 \vee 1/x - 2 > 0
    \vee y^2 < 5 \wedge 3 x^2 - 1/y
  + 1 < 0 ]],
$$
and we can be assured that the two formulas are fair-SAT
equivalent. 

Of
course, there are some rewritings that do not preserve fair-SATness.
For example, the axioms of equality imply
$\forall x[ \mydiv{1}{x} = \mydiv{1}{x} ]$.  Thus, we would expect
$\exists x[x=0]$ and $\exists x[x=0 \wedge \mydiv{1}{x} =
  \mydiv{1}{x}]$ to be equivalent.
However, the former is fair-SAT, while the latter is not.  In general,
rewritings that introduce new denominators or remove denominators have
the possibility to change the fair-SATness of a formula.
This appears to be unavoidable.  None the less, the situation with
fair-SAT seems to be much better than for usual computer algebra
practice, since there is at least a clear 
class of basic logical rewriting operators that preserve
fair-SATness.

\subsubsection{The definition of a well-defined formula}
The process of clearing denominators without worrying about whether
they might be zero is, of course, well known to anyone.  However, we
give a formal description (Algorithm~\ref{algorithm:clear}) of one
particular procedure so that we may reason about it in proofs and
definitions moving forward. Note that Algorithm~\ref{algorithm:clear}
assumes that there are no $\mydiv{\cdot}{\cdot}$ terms
nested in numerators.  This is easy to achieve via
$\mydiv{a}{b} = a\cdot\mydiv{1}{b}$ which, we note, does not affect
the fair-SATness of a formula.

\begin{algorithm}
\caption{clear$(A)$}
\label{algorithm:clear}
\SetKwInOut{Input}{input}\SetKwInOut{Output}{output}%
\Input{inequality $A$ with $\mydiv{\cdot}{\cdot}$ terms,
  where no numerator contains a nested $\mydiv{\cdot}{\cdot}$ term}
\Output{inequality $H$ without $\mydiv{\cdot}{\cdot}$ terms such that
  if $X$ is an $n$-tuple of variables that includes all the variables in
  $A$, for all $\gamma \in \realring^n$, if $A$ is legal at $X=\gamma$
  then $\pevalp{A}{X}{\gamma} = \pevalp{H}{X}{\gamma}$}
  let $A^* = A$ \;
  \While{nested $\mydiv{\cdot}{\cdot}$ terms remain in $A^*$}
  {
    choose a $\mydiv{\cdot}{\cdot}$ term $T$ of maximal nesting in
    $A^*$ [Note: $T = \mydiv{q}{p}$ where $q,p$ are div-free] \;
    note that $T$ is nested in some term $T' = \mydiv{d}{u + v T^j}$, replace
    $T'$ with
    $\mydiv{p^j d}{u p^j + v q^j}$ \;
  }
  \While{$\mydiv{\cdot}{\cdot}$ terms remain in $A^*$}
  {
    choose a $\mydiv{\cdot}{\cdot}$ term $T$ in  $A^*$
    [Note: $T = \mydiv{q}{p}$ where $q,p$ are div-free] \;
    Note: since $T$ is not nested,
    $A^*$ is of the form $u + v T^j\ \relop\ w$, where
    $\relop \in \{=,\neq,<,>,\leq,\geq\}$.\\
    replace $A$ with $u p^{j'} + v q^j p^{j'-j}\ \relop\ w p^{j'}$,
      where $j' = j+1$ if $j$ odd and $\relop \in \{<,>,\leq,\geq\}$,
      and $j' = j$ otherwise \;
  }
  $H = A^*$
\end{algorithm}

\begin{theorem}
  Algorithm~\ref{algorithm:clear} meets its specification.  
\end{theorem}

This theorem is an easy consequence of
Theorem~\ref{theorem:mainalgworks} (appearing later),
which proves the correctness of Algorithm~\ref{algorithm:translate},
an augmented 
version of Algorithm~\ref{algorithm:clear}.

\begin{definition}
  If $F$ is a formula, possibly with $\mydiv{\cdot}{\cdot}$ terms,
  $\clearf{F}$ is the formula resulting from replacing each
  atom $A$ in $F$ that contains $\mydiv{\cdot}{\cdot}$ terms with
  $\clearf{A}$.
\end{definition}

For formula $F$, $\clearf{F}$ is the result
of clearing denominators without adding any ``guards''.  
Conceptually, we say that a formula is ``well defined'' when this simple-minded
clearing of denominators is enough, since the formula is somehow
already doing the ``guarding''.  More formally:

\begin{definition}[Well defined formula]
  Formula $F$ with free variables $X = (x_1,\ldots,x_k)$, possibly with
  divisions, is \emph{well defined} if and only if for every
  $\gamma \in \realring^k$,
  $\pevalp{clear(F)}{X}{\gamma}$ is true if and only if
  $\pevalp{F}{X}{\gamma}$ is fair-SAT.
\end{definition}

\begin{example}
  Consider formula $F_1$ from Section~\ref{section:firstattempt}:
  $$
  F_1 := \exists x,y\left[y^2\left(1 + x^2 + \frac{1}{y^2}\right) \leq 0\right].
  $$
  Since there are no free variables, $F_1$ is well-defined if and only
  if $\clearf{F_1}$ is equivalent to fair-SAT $F_1$.
  $\clearf{F_1} = \exists x,y\left[y^6\left(y^4 + x^2 + 1\right) \leq 0\right]$
  which is true.
  However, $F_1$ is not fair-SAT, since for non-zero $\beta$ we have
  $\peval{F_1}{(x,y)}{(\alpha,\beta)} = \beta^2\left(1 + \alpha^2 + \frac{1}{\beta^2}\right) \leq 0$,
  which is false and thus not fair-SAT, and for $\beta=0$ we have
  $\peval{F_1}{(x,y)}{(\alpha,\beta)}=U_1$, which is not fair-SAT.
  Therefore, formula $F_1$ is not well defined.
\end{example}

\begin{example}
The next example illustrates the definition of well defined in the
presence of free variables.
Consider the formula
$F_3 := \exists y[y \geq a^2 \wedge (y^2 + 1)/y^2 \leq 1 + a]$,
and note that $a$ is free in $F_3$.  This formula is not fair-SAT,
which we see as follows.  For $F_3$ to be well-defined, $\peval{F_3}{a}{\gamma}$
must be well-defined for all real values $\gamma$.
Consider $\gamma = 0$.  Then $\peval{F_3}{a}{\gamma} = G$, where 
$G := \exists y[ y \geq 0 \wedge (y^2 + 1)/y \leq 1]$.
\begin{itemize}
\item
  $\peval{clear(F_3)}{a}{\gamma} = \exists y[y >= 0 \wedge y^2 (y^2 + 1) \leq y^4]$,
  which is $\true$.
\item $G$ is fair-SAT iff there is a $\beta$
  such that $H := \peval{[y \geq 0 \wedge (y^2 + 1)/y \leq 1]}{y}{\beta}$
  is fair-SAT.
  $$
  \begin{array}{l}
    \text{case $\beta < 0$: } H \text{ is } \false \text{ which is not fair-SAT}\\
    \text{case $\beta = 0$: } H \text{ is } \true\wedge U_1 \text{ which is not fair-SAT}\\
    \text{case $\beta > 0$: } H \text{ is } \true\wedge\false \text{ which is not fair-SAT}\\
  \end{array}
  $$
  So $G$ is not fair-SAT, and thus $\peval{F_3}{a}{\gamma}$ is not fair-SAT.
\end{itemize}
So ``$\pevalp{clear(F)}{X}{\gamma}$ is true if and only if
  $\pevalp{F}{X}{\gamma}$ is fair-SAT'' fails for $F_3$ at $a = \gamma = 0$, and
therefore $F_3$ is not well defined.
However, if we change the first constraint to $y > a^2$,
the resulting formula is well-defined.
\end{example}

\begin{example}
  Consider formula $F_4 := \forall x[ 1/x = 1/x]$.  
  clear$(F)$ is $\forall x[ x = x]$ which is true.
  $F_4$ is fair-SAT if and only if $F'_4$ := $\exists x[ 1/x \neq 1/x]$ is not
  fair-SAT.  $F'_4$ is not fair-SAT because for non-zero value $\alpha$
  $\peval{F'_4}{x}{\alpha}$ is false, which is not fair-SAT, and for
  $\alpha=0$,
  $\peval{F'_4}{x}{\alpha}$ is $U_1$, which is also not fair-SAT.
  This means $F_4$ is fair-SAT and, therefore, $F_4$ is well-defined.
\end{example}

\section{Handling formulas that are not well defined}
\label{section:guard}
Ideally, users or programs that call polynomial constraint solvers
with input formulas that include divisions would only send solvers
well-defined formulas, since we could just clear denominators without
having to worry about whether they vanish or not.
However, experience shows that this is not
always the case.  This motivates us to consider how best to handle
such formulas. Our goal is a translation procedure that maps formulas
with divisions into purely polynomial formulas such that:
1) a well-defined formula $F$ is mapped to polynomial formula
that is equivalent to the result of simply clearing denominators from $F$,
and
2) non-well-defined formula $F$ is mapped to a purely polynomial
formula that is SAT if and only if $F$ is fair-SAT.

\subsection{The translation algorithm}
In this section we present an algorithm called \texttt{translate}, which
translates formulas with divisions into purely polynomial formulas.
Our algorithm assumes that the given formula is prenex and
``positive''.

\begin{definition}
  We say formula $F$ is \emph{positive} provided that the only propositional
  operators appearing in $F$ are $\wedge,\vee,\neg$, and for any
  subformula of the form $\neg F'$, $F'$ is a propositional variable. 
\end{definition}

Although this restriction seems like a limitation, every formula has a
positive form that is easy to construct.

\begin{definition}
Let $F$ be a formula in which the only propositional operators are
$\wedge,\vee,\neg$. The \emph{positive form of $F$} is the result of
pushing the negation symbols ``inwards'' in the usual way until they
are either absorbed into inequalities or sit in front of a
propositional variable.
Note that there is
a natural correspondence between the atoms, quantifier blocks and $\wedge/\vee$ operators
of $F$ and its positive form.
\end{definition}

\begin{example}
  The positive form of $\neg ( x < y \vee (P \wedge y \leq 0) )$
  is $x \geq y \wedge (\neg P \vee y > 0)$.
  $$
  \begin{array}{rcccccccccc}
    F: & \neg ( & x < y & \vee & ( &  & P &  \wedge & y \leq 0 & ) & )\\
    \text{positive form of }F:&& x \geq y & \wedge & ( & \neg & P & \vee & y > 0& ) &
  \end{array}
  $$
\end{example}

Importantly, the rewritings used to produce the positive and prenex form of 
a given formula are ones that, as described in Section~\ref{section:fairSatProperties}
though proved below, preserve fair-SATness.  Thus, a formula and its
positive prenex form are fair-SAT equivalent, so requiring positive
prenex input to the translation procedure is not really a
limitation.

There is one more definition that we require in order to introduce our
translation procedure.  It involves a concept that is important for many
algorithms that do quantifier elimination and SAT-solving for elementary real algebra:
\emph{nullification} \cite{McCallum:97,McCallum:99,Brown:00b,McCallumEtAl:2019,NairDavenportSankaran:2020}.  Non-zero polynomial $p(x_1,\ldots,x_n)$ is said to be
\emph{nullified} at point $\alpha\in\realring^k$, where $k < n$, if
$p(\alpha_1,\ldots,\alpha_k,x_{k+1},\ldots,x_n)$ is zero, i.e. is the
zero element in $\realring[x_{k+1},\ldots,x_n]$.  This description
assumes the variable order $x_1 \prec \cdots \prec x_n$, but
nullification of a polynomial is defined more generally with respect
to a variable order, so $p = x z - y$ is nullified at $(0,0)$ with
respect to $x \prec y \prec z$, but nullified nowhere with respect to
$x \prec z \prec y$, since the $y$ term does not vanish at any point
in $(x,z)$-space.

It is not necessarily obvious that nullification has any connection to
fair-SATness or any of the other concepts considered thus far.  But
that connection is there: it comes from the dependence of partial
evaluation on whether denominators are \emph{legal}.  For example,
with variable order $x \prec y \prec z$, the term $\mydiv{1}{x z - y}$ is 
illegal at $(x,y) = (\alpha,\beta)$ exactly when $\alpha z - \beta$ is
zero for all values of $z$.  The fundamental theorem of algebra tells
us that this is only possible if $\alpha z - \beta$ is the zero
polynomial in $\realring[z]$, i.e. if
it is nullified at $(\alpha,\beta)$.

\begin{definition}  
  Let $p$ be a polynomial and let $V_1,\ldots,V_k$ be pair-wise
  non-intersecting sets of variables.
  The \emph{$i$-level nullifying system for $p$}, denoted
  $\nullsys{p}{i}$, is $c_1\wedge c_2\wedge \cdots \wedge c_t$, where
  the $c_i$s are the coefficients of $p$ as a polynomial in the
  variables $V_i \cup \cdots \cup V_k$.
  Note: if $i > k$ or $p$ has no variables in $V_i \cup \cdots \cup V_k$,
  $\nullsys{p}{i} = [p = 0]$.
\end{definition}

\begin{example}
  For example, consider the prenex formula
  $F := \exists x,y\forall z[x z^2 + 2 y z + x a \neq 0]$.
  Its quantifier blocks give us variable sets $V_1 = \{x,y\}$, $V_2 = \{z\}$.
  We examine the nullifying sets for polynomial
  $p = x z^2 + 2 y z + x a$.
  The coefficients of $p$ as a polynomial in $V_1\cup V_2 = \{x,y,z\}$ are $1,2,a$, so
  $\nullsys{p}{1} = [1=0 \wedge 2 = 0 \wedge a = 0]$.
  The coefficients of $p$ as a polynomial in $V_2 = \{z\}$ are $x, 2y, x a$, so
  $\nullsys{p}{2} = [x=0 \wedge 2 y = 0 \wedge x a = 0]$.
\end{example}


In the following, we will assume we start with formulas that are
prenex and positive, and in which any $\mydiv{\cdot}{\cdot}$
term $\mydiv{p}{q}$ has been rewritten as $p\cdot\mydiv{1}{q}$.

\begin{algorithm}[H]
\caption{translate$(F,A)$}
\label{algorithm:translate}
\SetKwInOut{Input}{input}\SetKwInOut{Output}{output}%
\Input{positive prenex formula $F$ (without $U_i/V_i$ variables) in which all div-terms are
of the form $\mydiv{1}{\cdot}$, where there are $k$ quantifier blocks $B_1,\ldots,B_k$,
each with variable sets $V_1,\ldots,V_k$, respectively;
(in)equality $A$ in $F$ with $\mydiv{\cdot}{\cdot}$ terms}
\Output{Formula $H_0$ (without division) such that
the following holds:
Let $x_{i_1}\prec\cdots\prec x_{i_n}$ be an ordering of the variables that
respects the quantifier block structure,
\begin{enumerate}[label=(\alph*)]
\item 
  if $\gamma\in\realring^t$, where $t \leq n$, is a
  point such that $A$ is legal at $(\gamma_1,\ldots,\gamma_{t-1})$ but
  not at $\gamma$,  then
  for all $\beta\in\realring^{n-t}$, 
  $\pevalp{H_0}{(x_{i_1},\ldots,x_{i_{t}})}{\gamma:\beta}$
  is $\true$ if $x_{t}$ is in a $\forall$-block, and $\false$
  otherwise,
  and
\item if $A$ is legal at some point $\gamma'\in\realring^n$, then
  $\pevalp{A}{(x_{i_1},\ldots,x_{i_{n}})}{\gamma'}
  \Leftrightarrow
  \pevalp{H_0}{(x_{i_1},\ldots,x_{i_{n}})}{\gamma'}$.
\end{enumerate}}
let $A^* = A$ \;
initialize each of $N_1,\ldots,N_{k+1}$ to the empty set \;
  \While{nested $\mydiv{\cdot}{\cdot}$ terms remain in $A^*$}
  {
    choose a $\mydiv{\cdot}{\cdot}$ term $T$ of maximal nesting in
    $A^*$ [Note: $T = \mydiv{q}{p}$ where $q,p$ are div-free] \;
    let $s$ be the maximum index for which $p$ has non-zero degree
    in some variable in $V_s$ \;
    for each $i\in\{1,\ldots,s+1\}$, add to set $N_i$
    the formula $\nullsys{p}{i}$ \;
    note that $T$ is nested in some term $T' = \mydiv{d}{u + v T^j}$, replace
    $T'$ with
    $\mydiv{p^j d}{u p^j + v q^j}$ \;
  }
  \While{$\mydiv{\cdot}{\cdot}$ terms remain in $A^*$}
  {
    choose a $\mydiv{\cdot}{\cdot}$ term $T$ in  $A^*$ [Note: $T =
      \mydiv{q}{p}$ where $q,p$ are div-free] \;
    let $s$ be the maximum index for which $p$ has non-zero degree
    in some variable in $V_s$ \;
    for each $i\in\{1,\ldots,s+1\}$, add to set $N_i$
    the formula $\nullsys{p}{i}$ \;
    Note: since $T$ is not nested,
    $A^*$ is of the form $u + v T^j\ \relop\ w$, where
    $\relop \in \{=,\neq,<,>,\leq,\geq\}$.\\
    replace $A^*$ with $u p^{j'} + v q^j p^{j'-j}\ \relop\ w p^{j'}$,
      where $j' = j+1$ if $j$ odd and $\relop \in \{<,>,\leq,\geq\}$,
      and $j' = j$ otherwise \;
  }
  let $H_{k+1} = A^*$ [note: $A^*$ has been rewritten to eliminate
  $\mydiv{\cdot}{\cdot}$ terms] \;
  let $C = \false$ \;
  \For{$i$ from 1 to $k+1$}
  {
    let $W = \bigvee_{K\in N_i} K$ \;
    let $G_i$ be $W$ or an appropriate simplification
    [i.e. $G_i$ satisfies $\neg C \Rightarrow (G_i \Leftrightarrow W)$] \;
    $C = C \vee G_i$ \;
  }
  \For{$i$ from $k+1$ down to 1}
  {
    if $i > 1$ and block $B_{i-1}$ is a $\forall$-block,
    then $H_{i-1} = G_i \vee H_i$,
    else $H_{i-1} = \neg G_i \wedge H_{i}$ \;
  }
  return $H_0$ \;
\end{algorithm}

\textbf{Note:}
If in Algorithm~\ref{algorithm:translate} we return
$H_{k+1}$, we get the exact same result as
Algorithm~\ref{algorithm:clear}.
So the proof that Algorithm~\ref{algorithm:translate} meets its
specification (Theorem~\ref{theorem:mainalgworks}), also proves that
Algorithm~\ref{algorithm:clear} meets its specification.
Also, if on line 14 we set $H_{k+1} = A$ rather than $H_{k+1} = A^*$,
Algorithm~\ref{algorithm:translate} can be viewed as adding ``guards''
for inequality $A$ with denominators, but without clearing denominators.
So, in essence, Algorithm~\ref{algorithm:translate} does two things:
it adds guards and it clears denominators.  But with trivial
modifications, it does either one individually.

\begin{example}
  We will step through translate$(F,A)$ on the following formula $F$,
  $A$ being its only atom:
  $$
  F :=
  \forall a, b
  \exists x [x - \mydiv{1}{a x + b\cdot \mydiv{1}{b+1}} =  0]
  $$
  so $B_1 = \forall a,b$ and $B_2 = \exists x$, so $k=2$ and
  $V_1=\{a,b\}$ and $V_2=\{x\}$. Going line-by-line:
  \begin{description}[noitemsep,topsep=3pt,parsep=3pt,partopsep=0pt]
  \item{lines 1-2:}
    $A^* := x - \mydiv{1}{a x + b\cdot \mydiv{1}{b+1}} =  0$,
    $N_1=\{\ \}$,$N_2=\{\ \}$,$N_3=\{\ \}$
  \item{line 4:} choose $T = \mydiv{1}{b+1}$
  \item{line 5:} $s = 1$
  \item{line 6:}
    $\nullsys{b+1}{1} = [1=0\wedge 1=0] = \false$,
    $\nullsys{b+1}{2} = [b + 1 = 0 ]$
  \item{line 7:} replace $\mydiv{1}{a x + b\cdot \mydiv{1}{b+1}}$
    with $\mydiv{b+1}{(b+1)a x + b}$
  \item{now:}
    $A^* := x - \mydiv{b+1}{(b+1)a x + b} =  0$,
    $N_1=\{\ \false \}$,$N_2=\{\ b+1=0 \}$,$N_3=\{\ \}$
  \item{line 9:} choose $T = \mydiv{b+1}{(b+1)a x + b}$
  \item{line 10:} $s = 2$
  \item{line 11:}
    $\nullsys{(b+1)a x + b}{1} = [\false]$,
    $\nullsys{(b+1)a x + b}{2} = [(b+1)a = 0 \wedge b = 0] = [a = 0 \wedge b = 0]$,
    $\nullsys{(b+1)a x + b}{3} = [(b+1)a x + b = 0]$,
  \item{line 13:} replace 
    $x - \mydiv{b+1}{(b+1)a x + b} =  0$ with
    $((b+1)a x + b)x - (b+1) =  0$
  \item{now:}
    $A^* = ((b+1)a x + b)x - (b+1) =  0$,
    $N_1 = \{\false\}$,
    $N_2 = \{b+1=0,a=0\wedge b=0\}$,
    $N_3 = \{(b+1) a x + b = 0\}$
  \item{line 14:} $H_3 = ((b+1)a x + b)x - (b+1) = 0$
  \item{line 15:} $C = \false$
  \item{lines}
    $
    \begin{array}{clll}
      16: & i = 1 & i = 2 & i = 3\\
      17: & W = \false & W = b + 1 = 0 \vee a = 0 \wedge b = 0 & W = (b+1)ax+b=0\\
      18: & G_1 = \false & G_2 = b + 1 = 0 \vee a = 0 \wedge b = 0 & G_3 = (b+1)ax+b=0\\
      19: & C = \false & C = b + 1 = 0 \vee a = 0 \wedge b = 0 &
      C = b + 1 = 0 \vee a = 0 \wedge b = 0 \vee (b+1)ax+b=0\\
    \end{array}
    $
  \item{lines}
    $
    \begin{array}{clll}
      20: & i = 3 & i = 2 & i = 1\\
      21: &
      H_2 = (b+1)ax+b=0 \neq 0 \wedge H_3 &
      H_1 = (b + 1 = 0 \vee a = 0 \wedge b = 0) \vee H_2 &
      H_0 = \neg \false \wedge H_1
    \end{array}
    $
  \item{return:} $H_0 =
    (b + 1 = 0 \vee a = 0 \wedge b = 0) \vee
    (b+1)ax+b \neq 0 \wedge ((b+1)a x + b)x - (b+1) = 0$
  \end{description}
  Note that at values for $(a,b)$ that make $A$ illegal, $H_0$ is true
  based on the left-hand disjunct. Since they are universally
  quantified, this matches the algorithm specification.
  Choosing values for $(a,b)$ at which $A$ is legal, but a value for
  $x$ that causes the denominator of the outer div to vanish results
  in $H_0$ being false based on the ``$\neq$'' constraint in the
  right-hand disjunct.  Since $x$ is existentially quantified, this
  matches the algorithm specification. Finally, for any other values
  of $(a,b,x)$, all of which are legal, the value of $H_0$ is determined by
  $((b+1)a x + b)x - (b+1) = 0$, which is the result of clearing
  denominators in $A$, and thus is equivalent to $A$ when no
  denominators vanish.

  Just looking at this example, in might appear that $N_1$ has no role
  to play and $\nullsys{p}{1}$ is always just $\false$.  However, this
  is because there are no free variables in the example.  When free
  variables are present, $N_1$ and $\nullsys{p}{1}$ have necessary
  roles to play.
\end{example}

The translation procedure for a formula $F$ is simple: just apply
translate$(F,A)$ to each atom that contains a div-term.

\begin{algorithm}[H]
\caption{translate$(F)$}
\label{algorithm:translateFormula}
\SetKwInOut{Input}{input}\SetKwInOut{Output}{output}%
\Input{positive prenex formula $F$ (without $U_i/V_i$ variables) potentially with divisions}
\Output{Formula $H$ (without division) with the 
same variables and quantifier blocks as $F$.
If $F$ has no free variables, then 
$H$ is SAT if and only if $F$ is fair-SAT.  If $F$ has free variables,
then if $X$ is the tuple of free variables, and $\Gamma$ a tuple of
values, $\pevalp{F}{X}{\Gamma}$ is fair-SAT if and only if
$\pevalp{H}{X}{\Gamma}$ is SAT.}
let $H = F$ \;
\While{there are div terms remaining in $H$}
{
  choose (in)equality $A$ with div terms \;
  let $H_0 = $ translate$(F,A)$ \;
  replace constraint $A$ with $(H_0)$ in $H$ \;
}
return $H$ \;
\end{algorithm}

\begin{example}
  Consider the formula $F := \forall a \exists b \forall x[x^2 + a x + 1/(a b) > 0]$.

  First we show that $F$ is fair-SAT.
  By definition, $F$ is fair-SAT if an only if
  $G := \exists a \forall b \exists x[x^2 + a x + 1/(a b) \leq 0]$
  is not fair-SAT. Let $G' := \forall b \exists x[x^2 + a x + 1/(a b) \leq 0]$,
  and consider $\peval{G'}{a}{\alpha}$.
  If $\alpha=0$, then one can
  verify that $\peval{G'}{a}{\alpha} = \forall b \exists x[U]$, where $U$ is a
  propositional variable, is not fair-SAT.
  If $\alpha \neq 0$, then
  $\peval{G'}{a}{\alpha} = \forall b \exists x[x^2 + \alpha x + 1/(\alpha b) \leq 0]$,
  and since there are always non-zero values for $\beta$ satisfying
  $\alpha^3 \beta - 4 < 0 \wedge \alpha \beta > 0$,
  and since such values result in a negative discriminant for $x^2 + \alpha x + 1/(\alpha b)$,
  this formula is not fair-SAT.  Thus, $G$ is not fair-SAT, which
  means $F$ \emph{is} fair-SAT.

  Simply clearing denominators produces 
  $F_1 := \forall a \exists b \forall x[(a b x^2 + a^2 b x + 1) (a b) > 0]$,
  which is $\false$.
  Applying the Translate algorithm to rewrite $F$ by rewriting its one and
  only atom produces
  $$F_2 := \forall a \exists b \forall x[a = 0 \vee b \neq 0 \wedge (a b x^2 + a^2 b x + 1) (a b) > 0],$$
  which is $\true$ as required (since $F$ is fair-SAT).

\end{example}


\begin{example}
  \label{ex:nullmatters}
  Consider the formula $F := \forall a,b[\exists x[ b^2 + 4 a < 0 \vee x = 1/(a x + b) ]]$.
  Clearing denominators in the usual way yields
  $F_1 := \forall a,b[\exists x[ b^2 + 4 a < 0 \vee x (a x + b) = 1
  ]]$, which is $\false$.
  Following the Translate algorithm produces
  $$
  F_2 := \forall a,b[\exists x[ b^2 + 4 a < 0 \vee [a = 0 \wedge b = 0
        \vee a x + b \neq 0 \wedge x (a x + b) = 1 ]]],$$ which is $\true$.
  Note that, in this example, the polynomial in the denominator is
  irreducible, but nullified at $a=0\wedge b=0$, which involves only
  the variables in the universally quantified block. Had we ignored
  the nullification, we would have gotten
  $$
  F_3 := \forall a,b[\exists x[ b^2 + 4 a < 0 \vee 
      a x + b \neq 0 \wedge x (a x + b) = 1 ]],$$ which is $\false$.
  Thus, this example illustrates why we need to track the 
  nullification of polynomials appearing in denominators.
  
  %
\end{example}

Correctness proofs for translate$(F,A)$ and translate$(F)$ appear
in Section~\ref{section:correctnessProofs}.

\subsection{Where the usual guarding process fails}
In Section~\ref{section:capractice} we saw several examples of undesirable properties
of the usual guarding procedure used by computer algebra systems
fails.
We noted that $\exists x[1/x^2 < 0]$ and $\forall x[1/x^2 \geq 0]$ are
both false under the usual guarding procedure.  However, under the new
procedure we get:
$$
\exists x[1/x^2 < 0] \addguards \exists x[ x^2 \neq 0 \wedge 1 < 0 ]
\Longleftrightarrow \false
$$
$$
\forall x[1/x^2 \geq 0] \addguards \forall x[ x^2 = 0 \vee 1 \geq 0 ]
\Longleftrightarrow \true.
$$
We also noted that $\forall x,y[ x = y \Rightarrow 1/x = 1/y]$
is false under the usual guarding procedure.  However, under the new
procedure we get:
$$
\forall x,y[ x = y \Rightarrow 1/x = 1/y]
\addguards
\forall x,y[  x = y \Rightarrow [[x = 0 \vee y = 0] \vee y = x]] \Longleftrightarrow \true.
$$
The other example from Section~\ref{section:capractice} involves
the distinction between
$(x/y)^2 \geq 0$ and $\neg[(x/y)^2 < 0]$, which is not really relevant
for the new guarding procedure, since it operates on the positive form
of the original formula, which would eliminate this distinction.

\section{Proving properties of fair-SAT-ness}
\label{section:property}
The goal of this section is to show that for basic logical rewritings of a
formula that are not specific to real algebra and that
do not add or remove denominators, fair-SATness is preserved.
We show that moving quantifiers (in particular, for converting a
formula to prenex), 
moving around logical negation operators according to the usual rules,
and absorbing logical negation into comparison operators all
preserve fair-SATness.

Note: for the rest of this section, we will always assume that we are
dealing with formulas in which $\wedge/\vee/\neg$ are the only boolean
operators, and in which variables are standardized apart.

\subsection{Properties of the $U_i$ and $V_i$ variables}
Partial evaluation introduces $U_i$ variables, and the ``flop''
operation changes $U_i$ variables to $V_i$ variables (and back).
In this section we prove some basic results concerning these special
variables and their role in determining whether a formula is
fair-SAT. The first result, and its corollaries, concern formulas
containing only $U_i$ and $V_i$ variables, and boolean constants.
Such formulas are the base case in the recursive definition of
fair-SAT.  These results show that, while the definition of fair-SAT
boils down in the base case to a quantified boolean formula, we
actually only need to consider one set of assignments to determine the
truth of that quantified boolean formula.

\begin{theorem}
  \label{theorem:assignvs}
  Let $F$ be a formula that is \george{}, and which contains no real
  variables.
  If $U_i$ appears in $F$ then
  $$
  \forall U_i[F] \Longleftrightarrow F|_{U_i = b_i}
  $$
  where $b_i$ is false if, in the positive form of $F$, the literal
  containing $U_i$ is positive, and true otherwise.
  If $V_i$ appears in $F$ then
  $$
  \exists V_i[F] \Longleftrightarrow F|_{V_i = c_i}
  $$
  where $c_i$ is true if, in the positive form of $F$, the literal
  containing $V_i$ is positive, and false otherwise.
\end{theorem}
\proof
This is an obvious consequence of the fact that each $U/V$ variable appears
exactly once.
\qed

\begin{corollary}
  \label{corollary:reorder}
  If $F$ satisfies the hypotheses of Theorem~\ref{theorem:assignvs},
  then the truth of 
  $\forall U_{i_1}\ldots U_{i_a} \exists V_{i_1}\ldots V_{i_b} [F]$
  is the same no matter how the quantifiers are reordered.
\end{corollary}

\begin{corollary}
  \label{corollary:flopbase}
  If $F$ satisfies the hypotheses of Theorem~\ref{theorem:assignvs},
  then $F$ is fair-SAT if and only if $\flop{F}$ is not fair-SAT.
\end{corollary}
\proof
We show that not $F$ fair-SAT is equivalent to $\flop{F}$ fair-SAT.
$$
\begin{array}{rcll}
\neg\forall U_{i_1}\ldots U_{i_a} \exists V_{i_1}\ldots V_{i_b} [F]
& \Longleftrightarrow &
\exists U_{i_1}\ldots U_{i_a} \forall V_{i_1}\ldots V_{i_b} [\neg F] &\\
 & \Longleftrightarrow &
\exists V_{i_1}\ldots V_{i_a} \forall U_{i_1}\ldots U_{i_b} [\flop{F}],
& \text{renaming $V$s as $U$s and vice versa}\\
 & \Longleftrightarrow &
\forall U_{i_1}\ldots U_{i_b} \exists V_{i_1}\ldots V_{i_a}  [\flop{F}],
& \text{by Corollary~\ref{corollary:reorder}.}\\
\end{array}
$$
\qed

The following theorem builds on Theorem~\ref{theorem:assignvs}
by showing that, as long as we are dealing with positive, prenex
formulas, replacing the $U_i$ and $V_i$ variables with
appropriately chosen boolean constants produces a fair-SAT equivalent
formula.  This theorem plays a crucial role in the proof-of-correctness
for the translate algorithm in Section~\ref{section:correctnessProofs}.

  \begin{theorem}
  \label{theorem:UsVs2BoolConsts}
  Let $F$ and $F'$ be prenex, positive and peval-safe formulas that
  differ only in that either:
  (a) $U_i$ occurs in $F$ but in $F'$ has been replaced with
  boolean constant $b_i$, where $b_i$ is false if the literal
  containing $U_i$ is positive, and true otherwise, or
  (b) $V_i$ occurs in $F$ but in $F'$ has been replaced with
  boolean constant $c_i$, where $c_i$ is true if the literal
  containing $V_i$ is positive, and false otherwise.
  $F$ and $F'$ are fair-SAT equivalent. 
  \end{theorem}
  \proof
  We will assume that $F$ (and thus $F'$) have no free (unquantified)
  real variables.  If not, we note that for any assignment of values
  to the free variables, the resulting partial evaluation of $F$ and
  $F'$ will yield formulas satisfying the conditions of the theorem
  with the same variable $U_i$ (respectively $V_i$)
  in the same literal in the ``new'' $F$, and with the same boolean
  constant in the associated position in the ``new'' $F'$.
  
  We prove the theorem by induction on $k$, the number of quantifier
  blocks in $F$. If $k = 0$, the theorem is equivalent to
  Theorem~\ref{theorem:assignvs}.  If $k > 0$, we consider two cases:

  Case 1: The outermost block consists
  of existentially quantified variables $x_j,\ldots,x_{j+d}$. By
  definition $F$ is fair-SAT if and only if there exist real constants
  $\alpha_{j},\ldots,\alpha_{j+d}$ such that
  $\peval{F_{qf}}{(x_j,\ldots,x_{j+d})}{(\alpha_{j},\ldots,\alpha_{j+d})}$
  is fair-SAT, where $F = \exists x_j,\ldots,x_{j+d}[F_{qf}]$.
  Similarly, by
  definition $F'$ is fair-SAT if and only if there exist real constants
  $\alpha_{j},\ldots,\alpha_{j+d}$ such that
  $\peval{F'_{qf}}{(x_j,\ldots,x_{j+d})}{(\alpha_{j},\ldots,\alpha_{j+d})}$
  is fair-SAT, where $F' = \exists x_j,\ldots,x_{j+d}[F'_{qf}]$.
  Since $\peval{F_{qf}}{(x_j,\ldots,x_{j+d})}{(\alpha_{j},\ldots,\alpha_{j+d})}$
  and
  $\peval{F'_{qf}}{(x_j,\ldots,x_{j+d})}{(\alpha_{j},\ldots,\alpha_{j+d})}$
  satisfy the requirements of the theorem, but with fewer quantifier
  blocks, we conclude by induction that they are fair-SAT equivalent.
  Thus $F$ and $F'$ are fair-SAT equivalent.

  Case 2: The outermost block consists
  of universally quantified variables $x_j,\ldots,x_{j+d}$.
  In this case, $F$ is fair-SAT if and only if
  $\overline{F} := \flop{F}$ is not fair-SAT. 
  Similarly, $F'$ is fair-SAT if and only if
  $\overline{F}' := \flop{F'}$ is not fair-SAT.

  Sub-case: Suppose $U_i$ appears positively in $F$. Then $\neg V_i$ appears
  at the same location in $\overline{F}$.
  In $F'$, the constant $\false$ appears in place of $U_i$,
  while in $\overline{F'}$, that $\false$ is replaced with
  $\neg\true$.
  So, $\overline{F}$ and $\overline{F'}$ are identical, other than
  $\true$ appearing in $\overline{F'}$ where $V_i$ appears
  in $\overline{F}$.
  This means they satisfy the hypotheses of the theorem, though still
  with $k$ quantifier blocks.
  However, their outermost quantifier blocks consist of existentially
  quantified variables $x_j,\ldots,x_{j+d}$, so we can revert to Case
  1 to finish the proof. 

  The other sub-cases:
  $U_i$ appears negatively in $F$,
  $V_i$ appears positively in $F$, and
  $V_i$ appears negatively in $F$ can be handled analogously.
  \qed

  Finally, we show that whether a $U_i$ variable appears positively or
  negatively does not really matter in determining whether a formula
  is fair-SAT. 
  
  \begin{lemma}
    \label{lemma:UiViNegation}
If closed formulas $F$ and $F'$ are both $\george$, with $\#F = \#F'$, and
which differ only in that for exactly one $i$ such that $1 \leq i \leq \#F$,
$U_{k_i}$ appears in one while $\neg U_{k_i}$ appears in the other, or
$V_{k_i}$ appears in one while $\neg V_{k_i}$ appears in the other, then
$F$ and $F'$ are fair-SAT equivalent.
\end{lemma}
\proof
We prove this by induction on the size of the formula defined as
follows. Quantifier-free formulas have size zero. Otherwise the
size is $1 + k_Q + k_b + 2^q \cdot k_n$, where $q$ is the number of
quantifiers in the formula and
\begin{itemize}[noitemsep,topsep=0pt,parsep=0pt,partopsep=0pt]
\item $k_Q$ is the number defined 
  by viewing the quantifiers appearing in the formula from left to right
  as a binary number, where $\forall = 1$ and $\exists = 0$,
\item $k_b$ is the number of $\wedge/\vee$ operators, and
\item $k_n$ is the sum over all $\neg$ operators of the number of
  $\wedge/\vee/\exists/\forall$ operators in the argument
  to which the negation is applied.  
\end{itemize}
We go through the cases from the definition of fair-SAT one by one.

Case 1, $F$ and $F'$ contain no real variables.  Note that this is the
base case.
If
$U_{k_i}$ appears in one and $\neg U_{k_i}$ appears in the other, then
$U_{k_i}$ is universally quantified and appears only once.  So if the
one formula holds for both values of $U_{k_i}$, the other holds for both
values of $\neg U_{k_i}$.  If
$V_{k_i}$ appears in one and $\neg V_{k_i}$ appears in the other, then
$V_{k_i}$ is existentially quantified and appears only once.  So if 
one formula holds for one value for $V_{k_i}$, the other formula holds
when $V_{k_i}$ is assigned the opposite value.

Case 2, $F$ is the negation of a smaller subformula, and we have
subcases we must consider.  In each, the theorem clearly follows by
induction as long as we can show that the size of formulas is reduced
so induction does indeed apply.
\begin{itemize}[noitemsep,topsep=0pt,parsep=0pt,partopsep=0pt]
\item $F$ is $\neg(A \wedge B)$, which is fair-SAT if and only if
  $\neg A \vee \neg B$ is fair-SAT. Induction applies since
  $k_n$ is reduced because the $\wedge$ is no longer part of the
  argument for any $\neg$ operator.
\item $F$ is $\neg(A \vee B)$, same argument as above.
\item $F$ is $\neg\neg F'$, which is fair-SAT if and only if $F'$ is
  fair-SAT. $F'$ has at least one quantifier, since we would otherwise
  be in Case 1, so $k_n$ is reduced and induction applies.
\item $F$ is $\neg \exists x[F']$, which is fair-SAT if and only if
  $\forall x[\neg F']$ is fair-SAT.  In this situation, $k_n$ is
  reduced by one, which reduces the overall size by $2^q$, and $k_b$ is
  unchanged, but $k_Q$ may increase. 
  However, for any formula $k_Q < 2^q$, so the overall size is
  decreased and induction applies.
\item $F$ is $\neg \forall x[F']$, which is fair-SAT if and only if
  $\exists x[\neg F']$ is fair-SAT.  The argument for this is the same
  as that for the preceding.
\end{itemize}

Case 3, $F = A \vee B$.  Then $U_{k_i}$ or $V_{k_i}$ appears either in
$A$ or $B$, but not both. So $F = A' \vee B'$ where $A$ and $A'$ are
either identical, or satisfy the hypotheses of this lemma, and so are
fair-SAT equivalent either by virtue of being identical, or by
induction (note that for both $A/A'$ and $B/B'$, $k_b$ is reduced,
$k_Q$ is not increased and $k_n$ is unchanged).
The same argument applies to $B$ and $B'$. so $F$ and $F'$ are
fair-SAT equivalent.

Case 4, $F = A \wedge B$. This follows the same argument as the
previous case.

Case 5, $F = \exists x[G]$. Then $F' = \exists x[G']$, and for any
$\alpha\in\realring$, $\peval{G}{x}{\alpha}$ and $\peval{G'}{x}{\alpha}$
satisfy the hypotheses of this lemma, and thus by induction (note that
$k_Q$ is reduced, but $k_b$ and $k_n$ are unchanged) are
fair-SAT equivalent.  This means $F$ and $F'$ are fair-SAT equivalent.

Case 6, $F = \forall x [G]$.  $F$ is fair-SAT if and only if
$\exists x[\flop{G}]$ is fair-SAT. $F'= \forall x [G']$ is fair-SAT if and only if
$\exists x[\flop{G'}]$ is fair-SAT.  Note that $\exists x[\flop{G}]$ and
$\exists x[\flop{G'}]$ satisfy the requirements of this lemma, and are
strictly smaller than $F$ and $F'$,
since $k_Q$ is reduced, $k_b$ is unchanged, and $k_n$ is zero after
flop (because all the negations get pushed inside of the
$\wedge/\vee/\exists/\forall$ operators).
Therefore, by induction
they are
fair-SAT equivalent, and so $F$ and $F'$ are fair-SAT equivalent.
\qed

\subsection{Preserving fair-SATness across rewritings involving negation}

A feature of ordered rings is that logical negation can be absorbed
into relational predicate expressions which, of course, we write using
the usual relational operators.  For example, instead of $\neg[x > 0]$
we may write $[x \leq 0]$.  When examining the usual practice of
computer algebra systems in handling division, we saw that such
rewritings could change the meaning of a formula.  We first show that
this is not the case for fair-SATness.  And then show that
fair-SATness is preserved, more generally, across the conversion of a
formula to its positive form.

\begin{theorem}
  If $F$ contains an occurrence of atom A := $[f\ \sigma\ g]$, and $F'$
  is the formula resulting from replacing the occurrence of $A$ with
  $A' := \neg[f\ \overline{\sigma}\ g]$, then $F$ and $F'$ are fair-SAT
  equivalent. 
\end{theorem}
\proof
To prove this, we can apply an inductive argument that follows the
structure of the fair-SAT definition.  Eventually, $A$ will evaluate
to $\true$ or to $\false$, in which case $A'$ will evaluate to
$\neg\false$ or to $\neg\true$, respectively, which is clearly equivalent.
This is straightforward except
that we may, in applying Point 5 of the definition of fair-SAT,
have atoms $A$ and $A'$ get replaced with $U_{k_i}$ and
$\neg U_{k_i}$, respectively.  However, if this
happens, Lemma~\ref{lemma:UiViNegation} finishes off the proof. 
\qed

\begin{theorem}
  Let $F$ be a closed formula, possibly with $U_i$/$V_i$s.
  Let $F' = \posform{F}$.  $F$ is fair-SAT if and only if $F'$ is
  fair-SAT. 
\end{theorem}
\proof
We proceed by induction on the size of $F$, and consider the following
cases, which match the definition of fair-SAT:
\begin{enumerate}
\item $F$ contains only propositional variables, true/false constants
  and boolean operators.
  In this case $F$ and $F'$ are equivalent boolean formulas, so
  $$
  \forall U_{i_1},\ldots,U_{i_a}\exists V_{j_1},\ldots,V_{j_b}[F]
  \Leftrightarrow  
  \forall U_{i_1},\ldots,U_{i_a}\exists V_{j_1},\ldots,V_{j_b}[F'],
  $$
  which means $F$ and $F'$ are fair-SAT equivalent.
  Note: This is the base case!
\item $F$ is $\neg G$, and we distinguish the following subcases:
  \begin{enumerate}
    \item $F = \neg(A \vee B)$.  In this case, $F$ is fair-SAT if and
      only if $\neg A \wedge \neg B$ is fair-SAT.  Note that
      $F' = \posform{\neg A} \wedge \posform{\neg B}$, which is
      $\posform{\neg A \wedge \neg B}$.
      The fair-SAT equivalence of these two comes from case~\ref{case:and}.
    \item $F = \neg(A \wedge B)$.  In this case, $F$ is fair-SAT if and
      only if $\neg A \vee \neg B$ is fair-SAT.  Note that
      $F' = \posform{\neg A} \vee \posform{\neg B}$, which is
      $\posform{\neg A \vee \neg B}$.
      The fair-SAT equivalence of these two comes from
      case~\ref{case:or}.
    \item $F = \neg \neg G$.  In this case, $F$ is  fair-SAT if and
      only if $G$ is fair-SAT.  $F' = \posform{G}$.  By induction, $G$
      is fair-SAT equivalent to $\posform{G}$, so $F$ and $F'$ are
      fair-SAT equivalent.
    \item $F = \neg \exists x[G]$.  $F$ is fair-SAT if and only if
      $\forall x[\neg G]$ is fair-SAT.
      $\posform{F} = \forall x[ \posform{\neg G}] = \posform{\forall x[ \neg G]}$.  By
      case~\ref{case:forall},  $\forall x[\neg G]$ and
      $\posform{\forall x[ \neg G]}$ are fair-SAT equivalent.
    \item $F = \neg \forall x[G]$.  $F$ is fair-SAT if and only if
      $\exists x[\neg G]$ is fair-SAT.
      $\posform{F} = \exists x[ \posform{\neg G}] = \posform{\exists x[ \neg G]}$.  By
      case~\ref{case:exists},  $\exists x[\neg G]$ and
      $\posform{\exists x[ \neg G]}$ are fair-SAT equivalent.      
  \end{enumerate}
\item \label{case:or}
  $F = A \vee B$.  $F$ is fair-SAT if and only if $A$ is fair-SAT or
  $B$ is fair-SAT.  $F' = \posform{A \vee B} = \posform{A} \vee \posform{B}$.
  By induction, $A$ is fair-SAT equivalent to $\posform{A}$, and
  $B$ is fair-SAT equivalent to $\posform{B}$.  Therefore, $F$ and
  $F'$ are fair-SAT equivalent.  
\item \label{case:and}
  $F = A \wedge B$.  $F$ is fair-SAT if and only if $A$ is fair-SAT and
  $B$ is fair-SAT.  $F' = \posform{A \wedge B} = \posform{A} \wedge \posform{B}$.
  By induction, $A$ is fair-SAT equivalent to $\posform{A}$, and
  $B$ is fair-SAT equivalent to $\posform{B}$.  Therefore, $F$ and
  $F'$ are fair-SAT equivalent.  
\item \label{case:exists}
  $F = \exists x[G]$.  $F$ is fair-SAT if and only if there exists a
  real value $\alpha$ for
  which $\peval{G}{x}{\alpha}$ is fair-SAT.
  $F' = \posform{F} = \exists x[\posform{G}]$, which is fair-SAT if
  and only if there exists a real value $\alpha$ for which
  $\peval{\posform{G}}{x}{\alpha}$ is fair-SAT.  Note that $G$ and
  $\posform{G}$ have the same atoms in their corresponding positions,
  except that, in some cases, $G$ may have $f\ \sigma\ g$ at a position
  at which $\posform{G}$ has $f\ \overline{\sigma}\ g$.  Let $G^*$ be
  the formula resulting from replacing each instance of such an
  inequality in
  $G$
  with $\neg[f\ \overline{\sigma}\ g]$.  Note that by the previous
  theorem, $G^*$ and $G$ are fair-SAT equivalent.  Note that
  $\posform{G} = \posform{G^*}$.  Since $G^*$ and $\posform{G^*}$ have
  the same atoms, for any real value $\alpha$,
  $\posform{\peval{G^*}{x}{\alpha}} = \peval{\posform{G^*}}{x}{\alpha}$ and
  so by induction $\peval{G^*}{x}{\alpha}$ is fair-SAT equivalent to 
  $\peval{\posform{G^*}}{x}{\alpha}$.  So $\peval{G}{x}{\alpha}$ is fair-SAT equivalent to 
  $\peval{\posform{G}}{x}{\alpha}$.  Therefore, $F$ and $F'$ are
  fair-SAT equivalent.
\item \label{case:forall}
  $F = \forall x[G]$.  $F$ is fair-SAT if and only if $\exists x[\flop{G}]$
  is not fair-SAT.  $F' = \posform{F} = \forall x[\posform{G}]$
  is fair-SAT if and only if $\exists x[\flop{\posform{G}}]$ 
  is not fair-SAT.
  \begin{lemma}
    $\flop{G} = \flop{\posform{G}}$, and
    $\flip{G} = \flip{\posform{G}}$
    (note: ``$=$'' here means syntactically identical).
  \end{lemma}

  This lemma is easily seen to be true, as flip and flop compute a
  posform as they go.

  Since $\exists x[\flop{G})]$ and $\exists x[\flop{\posform{G}}]$ are
  syntactically identical, they are fair-SAT equivalent.  
\end{enumerate}
\qed




\subsection{Fair-SATness and quantifiers}

The last class of rewritings we prove preserve fair-SATness consists
of the usual rewritings involving quantifiers, e.g.
rewriting $A \wedge \exists x[B]$ as $\exists x[A \wedge B]$ (assuming
$x$ is not free in $A$).  To help with these proofs, we introduce
``$\fsefull{F}{\alpha}$'', 
the ``fair-SAT-evaluation of formula $F$ at point $\alpha$''.

\begin{definition}
  \label{definition:fse}
  Let $F$ be a formula that is $\george$, with the further stipulation that
  no negated subformula contains quantifiers.  We will refer to the
  variables appearing in $F$ in left-to-right order as $x_{k_1},\ldots,x_{k_n}$.
  Given
  $\alpha\in\realring^n$,
  the
  \emph{fair-SAT evaluation of $F$ at $\alpha$}, denoted
  $\fsefull{F}{\alpha}$ (or $\fse{F}$ if it is clear
  what point $\alpha$ we are evaluating at), is the
  boolean formula defined by the following rules:
  (observe that this definition directly follows the definition of fair-SAT)
  \begin{enumerate}
    \item if $F$ contains only boolean constants and propositional
      variables,
      $$\fse{F} = \forall U_{i_1},\ldots,U_{i_a}
      \exists V_{j_1},\ldots,V_{j_b}[F],$$
      where
      $U_{i_1},\ldots,U_{i_a}$ are the $U$-variables and
      $V_{j_1},\ldots,V_{j_b}$ the $V$-variables in $F$.
      $\fse{F} = \neg\fse{\exists x_i[\flop{G}]}$\\
      Note: if this case applies it takes precedence over other cases.
    \item Note: if $F = \neg G$ the requirements of $F$ mean that we
      are in case 1.
    \item if $F = G \vee H$, then
      $\fse{F} = \fse{G} \vee \fse{H}$
    \item if $F = G \wedge H$, then
      $\fse{F} = \fse{G} \wedge \fse{H}$
    \item if $F = \exists x_{k_1}[G]$ then
      $\fsefull{F}{(\alpha_1,\ldots,\alpha_n)} = \fsefull{\peval{G}{x_{k_1}}{\alpha_1}}{(\alpha_2,\ldots,\alpha_n)}$
    \item if $F = \forall x_{k_1}[G]$ then
      $\fse{F} = \neg\fse{\exists x_{k_1}[\flop{G}]}$
  \end{enumerate}
\end{definition}

The mechanics of this definition, though perhaps not its utility, may
be made clearer with a concrete example.
Let $F := \forall x[\exists y[ y - 1/x^2 > 0 \vee x/(y-1) \neq 0]]$.
We illustrate $\fse{F}$ at four different points.
$$
\begin{array}{ll}
  \begin{array}{l}
    \fsefull{F}{(2,3)}\\
    =
    \neg\fsefull{\exists x[\forall y[ y - 1/x^2 \leq 0 \wedge x/(y-1) = 0]]}{(2,3)}
    \\
    =
    \neg\fsefull{\forall y[ y - 1/4 \leq 0 \wedge 2/(y-1) = 0]}{(3)}
    \\
    =
    \fsefull{\exists y[ y - 1/4 > 0 \vee 2/(y-1) \neq 0]}{(3)}
    \\
    =
    \fsefull{\true \vee \true}{}
    \\
    =
    \true \vee \true
  \end{array}
  &
  \begin{array}{l}
    \fsefull{F}{(0,3)}\\
    =
    \neg\fsefull{\exists x[\forall y[ y - 1/x^2 \leq 0 \wedge x/(y-1) = 0]]}{(0,3)}
    \\
    =
    \neg\fsefull{\forall y[ U_1 \wedge 0/(y-1) = 0]}{(3)}
    \\
    =
    \fsefull{\exists y[ \neg V_1 \vee 0/(y-1) \neq 0]}{(3)}
    \\
    =
    \fsefull{\neg V_1 \vee \false}{}
    \\
    =
    \exists V_1[ \neg V_1 \vee \false]
  \end{array}
\end{array}
$$
$$
\begin{array}{ll}
  \begin{array}{l}
    \fsefull{F}{(2,1)}\\
    =
    \neg\fsefull{\exists x[\forall y[ y - 1/x^2 \leq 0 \wedge x/(y-1) = 0]]}{(2,1)}
    \\
    =
    \neg\fsefull{\forall y[ y - 1/4 \leq 0 \wedge 2/(y-1) = 0]}{(1)}
    \\
    =
    \fsefull{\exists y[ y - 1/4 > 0 \vee 2/(y-1) \neq 0]}{(1)}
    \\
    =
    \fsefull{\true \vee U_1}{}
    \\
    =
    \forall U_1[ \true \vee U_1]
  \end{array}
  &
  \begin{array}{l}
    \fsefull{F}{(0,1)}\\
    =
    \neg\fsefull{\exists x[\forall y[ y - 1/x^2 \leq 0 \wedge x/(y-1) = 0]]}{(0,1)}
    \\
    =
    \neg\fsefull{\forall y[ U_1 \wedge 0/(y-1) = 0]}{(1)}
    \\
    =
    \fsefull{\exists y[ \neg V_1 \vee 0/(y-1) \neq 0]}{(1)}
    \\
    =
    \fsefull{\neg V_1 \vee U_2}{}
    \\
    =
    \exists V_1 \forall U_2[ \neg V_1 \vee U_2]
  \end{array}
\end{array}
$$

The purpose of defining $\fse{F}$ is that it gives us a different way
to talk about formulas being fair-SAT, which is made clear by the
following theorem.  Subsequently, we will leverage this new way to
characterize fair-SAT in order to prove that the usual rewritings
involving quantifiers preserve fair-SATness.

\begin{theorem}
  \label{theorem:fsfse}
  Let formula $F$ satisfy the
  requirements of Definition~\ref{definition:fse}.
  Let $x_{k_1},\ldots,x_{k_n}$ be the variables in $F$ in the
  left-to-right order they are introduced, with respective
  quantifiers $Q_{1},\ldots,Q_{n}$.
  $F$ is fair-SAT if and only if
  $Q_{1} \alpha_{1} Q_{2} \alpha_{2} \cdots Q_{n} \alpha_{n} [ 
    \fsefull{F}{(\alpha_1,\ldots,\alpha_n)}]$.    
\end{theorem}
\proof
We proceed by induction on $k = \sum_{i}^n \left(\text{1 if $Q_i=\exists$ and 2 otherwise}\right)$.
When $k = 0$ we are in case 1 of the defintion of $\fse{}$, which is
the same as the definition of fair-SAT, so the theorem holds.  Assume
$k > 0$.

If $F = A \vee B$, then $F$ is fair-SAT if and only if $A$ is fair-SAT
or $B$ is fair-SAT.
By definition, $\fse{F} = \fse{A} \vee \fse{B}$.
Note that for some $r$, $x_{k_1},\ldots,x_{k_r}$ are the variables appearing
in $A$, and $x_{k_{r+1}},\ldots,x_{k_n}$ are the variables appearing in $B$,
so
$
Q_{1} \alpha_{1} Q_{2} \alpha_{2} \cdots Q_{n} \alpha_{n} [\fsefull{F}{(\alpha_1,\ldots,\alpha_n)}]
\Leftrightarrow
Q_{1} \alpha_{1} \cdots Q_{r} \alpha_{r} [ \fsefull{A}{(\alpha_1,\ldots,\alpha_r)}]
\vee
Q_{r+1} \alpha_{r+1} \cdots Q_{k} \alpha_{k} [ \fsefull{B}{(\alpha_{r+1},\ldots,\alpha_k)}]$.
By induction, $A$ is fair-SAT if and only if
$Q_{1} \alpha_{1} \cdots Q_{r} \alpha_{r} [ \fsefull{A}{(\alpha_1,\ldots,\alpha_r)}]$
and $B$ is fair-SAT if and only if
$Q_{r+1} \alpha_{r+1} \cdots Q_{n} \alpha_{n} [ \fsefull{B}{(\alpha_{r+1},\ldots,\alpha_n)}]$,
so the theorem holds in this case.  The argument for the
$F = A \wedge B$ is basically the same.

If $F = \exists x_{k_1}[G]$, then $F$ is fair-SAT if and only if there
exists an $\beta\in\realring$ such that $\peval{G}{x_{k_1}}{\beta}$
is fair-SAT.  Note that $Q_1 = \exists$.  Assume $F$ is fair-SAT with
$\beta$ being a satisfactory value for $x_{k_1}$,
then $\peval{G}{x_{k_1}}{\beta}$ is fair-SAT, and by induction
the formula $Q_{2} \alpha_{2} \cdots Q_{n} \alpha_{n}\fsefull{\peval{G}{x_{k_1}}{\beta}}{(\alpha_2,\ldots,\alpha_n)}$
is true.  Since
$\fsefull{\peval{G}{x_{k_1}}{\beta}}{(\alpha_2,\ldots,\alpha_n)} = \fsefull{F}{(\beta,\alpha_2,\ldots,\alpha_n)}$,
we get that
$\exists \beta Q_{2} \alpha_{2} \cdots Q_{n} \alpha_{n}[\fsefull{G}{(\beta,\alpha_2,\ldots,\alpha_n)}]$ is true,
which is what we had to prove.
Starting with $\exists \beta Q_{2} \alpha_{2} \cdots Q_{n} \alpha_{n}[\fsefull{G}{(\beta,\alpha_2,\ldots,\alpha_n)}]$
being true, we can reverse this argument to show that $F$ is fair-SAT.

If $F = \forall x_{k_1}[G]$, then $F$ is fair-SAT if and only if 
$\exists x_{k_1}[\flop{G}]$ is not fair-SAT, which is equivalent to
the assertion that for any $\beta\in\realring$, $\peval{\flop{G}}{x_{k_1}}{\beta}$ is not
fair-SAT.
By induction, 
$\peval{\flop{G}}{x_{k_1}}{\beta}$ is not fair-SAT
if and only if $\neg[\overline{Q}_{2} \alpha_{2} \cdots
  \overline{Q}_{n}
  \alpha_{n}[\fsefull{\peval{\flop{G}}{x_{k_1}}{\beta}}{(\alpha_2,\ldots,\alpha_n)}]]$,
which we can rewrite as
$Q_{2} \alpha_{2} \cdots Q_{n} \alpha_{n}[\neg\fsefull{\peval{\flop{G}}{x_{k_1}}{\beta}}{(\alpha_2,\ldots,\alpha_n)}]$,
which is equivalent to 
$$Q_{2} \alpha_{2} \cdots Q_{n} \alpha_{n}[\neg\fsefull{\exists x_{k_1}\flop{G}}{(\beta,\alpha_2,\ldots,\alpha_n)}],$$
which is equivalent to
$$Q_{2} \alpha_{2} \cdots Q_{n} \alpha_{n}[\fsefull{F}{(\beta,\alpha_2,\ldots,\alpha_n)}]],$$
The assertion that this holds for any $\beta$ is 
$$\forall \beta Q_{2} \alpha_{2} \cdots Q_{n} \alpha_{n}[\fsefull{F}{(\beta,\alpha_2,\ldots,\alpha_n)}]],$$
  which is what we had to prove.
  \qed
  
  \begin{corollary}
    \label{mainfsfse}
    Let $F$ and $G$ be two formulas satisfying the requirements of
    Theorem~\ref{theorem:fsfse}, with the same variables introduced in
    the same order with the same quantifiers.
    If for all $\alpha\in\realring^n$
    the quantified boolean formulas $\fsefull{F}{\alpha}$ and
    $\fsefull{G}{\alpha}$ are equivalent, then $F$ and $G$ are fair-SAT equivalent.
  \end{corollary}

  Now we start proving that the most fundamental rewritings we expect for
  first-order logic preserve fair-SATness.

  \begin{theorem}
    If formula $\exists x_1[G] \wedge H$ satisfies the requirements of
    Theorem~\ref{theorem:fsfse}, then
    $\exists x_1[G] \wedge H$ and $\exists x_1[G \wedge H]$ are fair-SAT
    equivalent.    
  \end{theorem}
  \proof
  Noting that $H$ contains no occurrences of $x_1$:
  $$
  \begin{array}{rcl}
  \fsefull{\exists x_1[G] \wedge H}{\alpha}
  &=& 
  \fsefull{\exists x_1[G]}{(\alpha_1,\ldots,\alpha_r)} \wedge \fsefull{H}{(\alpha_{r+1},\ldots,x_n)}\\
  &=&
  \fsefull{\peval{G}{x_1}{\alpha_1}}{(\alpha_2,\ldots,\alpha_r)} \wedge \fsefull{H}{(\alpha_{r+1},\ldots,x_n)}\\
  \end{array}
  $$
  $$
  \begin{array}{rcl}
  \fsefull{\exists x_1[G \wedge H]}{\alpha}
  &=& 
  \fsefull{\peval{G \wedge H}{x_1}{\alpha_1}}{(\alpha_2,\ldots,x_n)}\\
  &=& 
  \fsefull{\peval{G}{x_1}{\alpha_1}  \wedge H}{(\alpha_2,\ldots,x_n)}\\
  &=&
  \fsefull{\peval{G}{x_1}{\alpha_1}}{(\alpha_2,\ldots,\alpha_r)} \wedge \fsefull{H}{(\alpha_{r+1},\ldots,x_n)}\\
  \end{array}
  $$
  \qed

  \begin{theorem}
    $\forall x[G] \wedge H$ and $\forall x[G \wedge H]$ are fair-SAT
    equivalent.
  \end{theorem}
  \proof
  $$
  \begin{array}{rcl}
  \fsefull{\forall x_1[G] \wedge H}{\alpha}
  &=&
  \fsefull{\forall x_1[G]}{(\alpha_1,\ldots,\alpha_r)} \wedge \fsefull{H}{(\alpha_{r+1},\ldots,\alpha_n)}\\
  &=&
  \neg\fsefull{\exists x_1[\flop{G}]}{(\alpha_1,\ldots,\alpha_r)} \wedge \fsefull{H}{(\alpha_{r+1},\ldots,\alpha_n)}\\
  \end{array}
  $$

  $$
  \begin{array}{rcl}
  \fsefull{\forall x_1[G \wedge H]}{\alpha}
  &=&
  \neg\fsefull{\exists x_1[\flop{G\wedge H}]}{(\alpha_1,\ldots,\alpha_n)}\\
  &=&
  \neg\fsefull{\exists x_1[\flop{G} \vee \flop{H}]}{(\alpha_1,\ldots,\alpha_n)}\\
  &=&
  \neg\left[
    \fsefull{\exists x_1[\flop{G}]}{(\alpha_1,\ldots,\alpha_r)}
    \vee
    \fsefull{\flop{H}}{(\alpha_{r+1},\ldots,\alpha_n)}
    \right]\\
  &=&
  \neg\fsefull{\exists x_1[\flop{G}]}{(\alpha_1,\ldots,\alpha_r)}
    \wedge
    \neg\fsefull{\flop{H}}{(\alpha_{r+1},\ldots,\alpha_n)}]\\
  \end{array}
  $$
\qed

%
%

\begin{theorem}
  Assume $H$ satisfies the conditions of Definition~\ref{definition:fse}.
$\fsefull{H}{\alpha} \Leftrightarrow \neg\fsefull{\flop{H}}{\alpha}$
\end{theorem}
\proof

Consider the case that $H$ has no real variables.  The theorem holds
in this case as a consequence of Corollary~\ref{corollary:flopbase}.

Consider the case $H = A\ op\ B$, where $op$ is $\wedge$ or $\vee$.
Then $\fse{H} = \fse{A}\ op\ \fse{B}$,
$\flop{H} = \flop{A}\ \overline{op}\ \flop{B}$, and
$\fse{\flop{H}} = \fse{\flop{A}}\ \overline{op}\ \fse{\flop{B}}$.
By induction, $\fse{A} \Leftrightarrow \neg\fse{flop{A}}$ and
$\fse{B} \Leftrightarrow \neg\fse{flop{B}}$.
So
$$
\begin{array}{rcl}
\neg\fse{\flop{H}} &=& \neg\fse{\flop{A}}\ op\ \neg\fse{\flop{B}}\\
&\Leftrightarrow & \fse{A}\ op\ \fse{B}.
\end{array}
$$

Consider the case that $H = \forall x_{k_1}[G]$.
By definition $\fse{H} = \neg\fse{\exists x_{k_1}[\flop{G}]}$, which is
exactly $\neg\fse{\flop{G}}$.

Finally, consider the case $H = \exists x_{k_1}[G]$. By definition,
$$
\fsefull{H}{\alpha} =
\fse{\peval{G}{x_{k_1}}{\alpha_1}}{(\alpha_2,\ldots,\alpha_n)}.
$$
$$
\neg\fsefull{\flop{H}}{\alpha} =
\neg\fsefull{\forall x_{k_1}[\flop{G}]}{\alpha} =
\neg\neg\fsefull{\exists x_{k_1}[\flop{\flop{G}}]}{\alpha}
= \fse{\peval{\flop{\flop{G}}}{x_{k_1}}{\alpha_1}}{(\alpha_2,\ldots,\alpha_n)}.
$$
\qed

\begin{corollary}
  If $\forall x[ G ]$ and $\exists x[ G]$ are \george{}, with $\#F = 0$
  (i.e. with no $U_i/V_i$ variables), then
  $\forall x[ G ]$ and $\neg \exists x[ \neg G]$ are fair-SAT equivalent,
  and
  $\exists x[ G ]$ and $\neg \forall x[ \neg G]$ are fair-SAT equivalent.
\end{corollary}

The above lemmas, theorems and colloraries combine to show that the
rewritings described in Section~\ref{section:fairSatProperties}
preserve fair-SATness.  In particular, we see that a formula $F$ and
is positive prenex form are fair-SAT equivalent, which is important,
because it allows us to apply the translation algorithm to arbitrary
input by first converting to positive prenex form.

\section{Correctness of the translation procedure}
\label{section:correctnessProofs}
\begin{theorem}
  \label{theorem:mainalgworks}
  Algorithm translate($F$,$A$) meets it specification.
\end{theorem}
\proof
First note that in every $\mydiv{\cdot}{\cdot}$ term in $A^*$ we
maintain the property that the numerators do not contain
$\mydiv{\cdot}{\cdot}$ terms as we make the rewriting steps on lines 7
and 13.

Let $I = \{ \gamma\in\realring^n|A^*\text{ illegal at }\gamma \}$.
and $N = \{ \gamma\in\realring^n|\gamma \text{ satisfies
}\bigvee_{i=1}^{k+1}\bigvee_{G\in N_i} G \}$, and consider the
conditions:
\begin{enumerate}[noitemsep,topsep=3pt,parsep=3pt,partopsep=0pt]
\item
  \label{item:invariant1}
  $I \cup N = \{ \gamma\in\realring^n|A\text{ illegal at }\gamma \}$.
\item 
  If $\mydiv{q'}{p'}$ is an occurrence of a div in $A^*$ and
  $\mydiv{1}{p}$ the associated occurrence of a div in $A$,
  then for any $\gamma\in\realring^n$ such that $\gamma \notin N$,
  $\peval{\mydiv{q'}{p'}}{(x_1,\ldots,x_n)}{\gamma}
  = \peval{\mydiv{1}{p}}{(x_1,\ldots,x_n)}{\gamma}$.
\item
  If $A$ is legal at $\gamma\in\realring^n$ then 
  $\peval{A}{(x_1,\ldots,x_n)}{\gamma}
  = \peval{A^*}{(x_1,\ldots,x_n)}{\gamma}$.
\end{enumerate}

These three conditions are loop invariant for both
``while'' loops of translate($F$,$A$).
They hold initially, when $A = A^*$ and taking them all
together, it's clear that they are maintained at each step of both
loops. By the time we get to line 14, $A^*$ has no $\mydiv{\cdot}{\cdot}$
terms, which means $N$ is exactly equal to the points at which
$A$ is illegal, and $A^*$ is illegal nowhere.

If $A$ is legal at some point $\gamma'\in\realring^n$, then since loop
invariant (1) tells us that, after the two while
loops, $N$ is exactly the illegal points of $A$, none of the formulas
in any of the $N_i$'s are satisfied at $\gamma'$.
Therefore, we see from the way $H_0$ is constructed that the truth of
$H_0$ at $\gamma'$ is equivalent to the truth of $H_{k+1}$, which is $A^*$.  Thus,
point (b) of translate$(F,S)$'s specification is
satisfied.\footnote{This also proves the correctness of Algorithm
clear$(A)$.}
So what remains is to prove point (a) of the specification.

Consider the div-terms of $A$: $\mydiv{q_1}{p_1},\ldots,\mydiv{q_m}{p_m}$.
Note that some $p_i$ may contain div-terms.  As the two while loops
are executed, the numerators and denominators of div-terms may change.
For each, however, before being eliminated they take on a final form
in which the denominator (and numerator) is a polynomial.  Denote these final forms of
$q_i$ and $p_i$ as $q^*_i$ and $p^*_i$.  The partial evaluation of
$\mydiv{q_i}{p_i}$ at a given point in $\realring^n$ is not $\fail$ 
if and only none of the div-terms nested within it peval to $\fail$ 
and the partial evaluation of $p_i$ is not zero.  Inductively, this
means that, if $\mydiv{q_{i_1}}{p_{i_1}},\ldots,\mydiv{q_{i_r}}{p_{i_r}}$
are the div-terms nested in $\mydiv{q_i}{p_i}$, then the partial
evaluation of
$\mydiv{q_i}{p_i}$ at a given point in $\realring^n$ is not $\fail$ 
if and only none of the polynomials
$$
p^*_i,p^*_{i_1},\ldots,p^*_{i_r}
$$
are zero at the given point.

If $\mydiv{q_i}{p_i}$ is illegal at a point in $\rho\in\realring^k$, where
$k < n$, then $\mydiv{q_i}{p_i}$ evaluates to $\fail$ for every point
that extends $\rho$ to dimension $n$.  This means that union of the
zero sets of
$$
p^*_i(\rho_1,\ldots,\rho_k,x_{k+1},\ldots,x_n),
p^*_{i_1}(\rho_1,\ldots,\rho_k,x_{k+1},\ldots,x_n), \ldots,
p^*_{i_r}(\rho_1,\ldots,\rho_k,x_{k+1},\ldots,x_n)
$$
must be the whole of $\realring^{n-k}$.  If a finite set of
$j$-variate polynomials with real coefficients has the property that the union of
their zero sets is $\realring^j$, then one of the polynomials in the
set must be the zero polynomial.  In our case, this means one of
$p^*_i,p^*_{i_1},\ldots,p^*_{i_r}$ must be nullified at $\rho$.

Finally, we turn to point (a) of the specification for translate$(F,A)$.
Suppose $\gamma\in\realring^t$, where $t \leq n$, is a
point such that $A$ is legal at $(\gamma_1,\ldots,\gamma_{t-1})$ but
not at $\gamma$.
Let $B_{j-1}$ be the quantifier block containing $x_t$.
Since none of the div-terms in $A$ are illegal
at $(\gamma_1,\ldots,\gamma_{t-1})$,
none of the $p^*_i$ polynomials are nullified at
$(\gamma_1,\ldots,\gamma_{t-1})$, and thus 
none of the formulas in any of
$N_1,\ldots,N_{j-1}$ are satisfied by
$(\gamma_1,\ldots,\gamma_{t-1})$, which means none of
$G_1,\ldots,G_{j-1}$ are satisfied at $(\gamma_1,\ldots,\gamma_{t-1})$.
Note that, for all $i$,  $G_i$ contains only free variables and variables from
quantifier blocks before $B_i$, so no extension of 
$(\gamma_1,\ldots,\gamma_{t-1})$ satisfies any of the $G_1,\ldots,G_{j-1}$.

Since some div-term is illegal at $\gamma$, some $p^*$ must be
nullified at $\gamma$, which means at least one of the formulas in $N_j$ is
satisfied at $\gamma$.  Consider the iteration $i = j$ of the for-loop
from lines 16--19.  Since $N_j$ is satisfied at $\gamma$, this
iteration's formula $W$ is satisfied at $\gamma$.
At line 18, $C = G_1 \vee \cdots G_{j-1}$,
which is false at $\gamma$ (from previous paragraph) so
$G_i \Leftrightarrow W$, which means $G_i$ is satisfied at $\gamma$.

Finally, consider line 21.  If $B_{j-1}$ is a $\forall$-block,
$H_{j-1} = G_j \vee H_j$, so $H_{j-1}$ is satisfied at any extension
of $\gamma$ to $\realring^n$ and,
since none of $G_1,\ldots,G_{j-1}$ are satisfied at $\gamma$, that
means $H_0$ is $\true$ at any extension of $\gamma$ to $\realring^n$.
If $B_{j-1}$ is not a $\forall$-block,
$H_{j-1} = \neg G_j \wedge H_j$, so $H_{j-1}$ is $\false$ at any
extension of $\gamma$ to $\realring^n$ and,
since none of $G_1,\ldots,G_{j-1}$ are satisfied at $\gamma$, that
means $H_0$ is $\false$ at any extension of $\gamma$ to $\realring^n$.
\qed

\begin{lemma}
  \label{lemma:flop2neg}
  If $F$ is a closed, positive, prenex formula, without $U_i$ or
  $V_i$ variables, then 
  $\text{translate}(\flop{F})
  \Leftrightarrow \neg
  \text{translate}(F)$.
\end{lemma}
\proof
Consider inequality $A$ in $F$ and
$H_0 = \text{translate}(F,A)$, which occupies the same
position in $H$ as $A$ does in $F$.  In $\flop{F}$, inequality $A$
is replaced with $\flop{A} = [f\ \overline{\sigma}\ g]$,
where $A = [f\ \sigma\ g]$
and, of course, the types of each quantifier block is toggled to the
other quantifier type.

We examine the difference between the execution of
$\text{translate}(F,A)$ and $\text{translate}(\flop{F},\flop{A})$.
Everything involved in the construction of the $G_i$s (lines 2--19)
depends only on $f$, $g$ and which variables occur in which blocks, not
the quantifier types of those blocks.
In Step 13, the only step that depends on the relational operator,
``$j'$'' depends on the relational operator,
 but in the same way for operator $\sigma$ as for $\overline{\sigma}$.
So in both executions, the
same $G_i$s are computed.  Using this, we will prove (below) that
the $H_{0}$ returned by
$\text{translate}(\flop{F},\flop{A})$ is equivalent to the negation of
the $H_{0}$ returned by $\text{translate}(F,A)$.  This completes the
proof because, while the translate algorithm replaces inequalities with
subformulas, it otherwise leaves the input formula unchanged. 

To facilitate induction,
we prove the slightly more general result that for any $i$,
 the $H_{i}$ produced while executing
 $\text{translate}(\flop{F},\flop{A})$ is equivalent to 
 the negation of the $H_{i}$ produced while executing $\text{translate}(F,A)$.
 To avoid unnecessary confusion, we will use a superscript $f$ to denote
 values computed by the execution of $\text{translate}(\flop{F},\flop{A})$
 as opposed to $\text{translate}(F,A)$.

 Our base case is $i = k+1$. $H^f_{k+1}$ is the negation of the
 $H_{k+1}$, since the construction
 of $A^*$ (lines 1--13) does not change the relational operator,
 or depend on it in
 any way other than in Step 13, in which ``$j'$'' depends on the
 relational operator,
 but in the same way for operator $\sigma$ as for $\overline{\sigma}$.
 So we consider the case $i < k+1$.
 
 Assume block $B_{i-1}$ is a $\forall$-block in $F$.
 Then $H_{i-1}= G_i \vee H_i$ and
 $H^f_{i-1}= \neg G_i \wedge H^f_i$, recalling that $G^f_i = G_i$.
 $\neg H_{i-1} = \neg G_i \wedge \neg{H_i}$ and by induction $\neg H_i$ is
 equivalent to  $H^f_i$, so $H^f_{i-1} = \neg H_{i-1}$.

 Assume block $B_{i-1}$ is an $\exists$-block in $F$.
 Then $H_{i-1}= \neg G_i \wedge H_i$ and
 $H^f_{i-1} = G_i \vee H^f_i$, recalling that $G^f_i = G_i$.
 $\neg H_{i-1} = G_i \vee \neg{H_i}$ and by induction $\neg H_i$ is
 equivalent to  $H^f_i$, so $H^f_{i-1} = \neg H_{i-1}$.
 \qed

\begin{theorem}
Algorithm translate$(F)$ meets its specification.
\end{theorem}
\proof
We prove this by induction on $2n + \delta$, where $n$ is the number
of real variables in $F$ and $\delta = 1$ if the outermost quantifier
is $\forall$ and $\delta = 0$ otherwise.
If $n=0$, there are no $\mydiv{\cdot}{\cdot}$ terms, and the
formula $H$ returned by translate$(F)$ is just $F$. With no
$\mydiv{\cdot}{\cdot}$ terms, there is no distinction between
fair-SAT and simply being $\true$, so the algorithm meets its
specification in this case.  So, for the remainder of the proof,
assume $n > 0$.

Consider first the case in which there are free variables. Let
$x := (x_j,\ldots,x_{j+d})$ be the free variables, 
$\alpha:=(\alpha_j,\ldots,\alpha_{j+d})$ be real values, 
$F^* := \peval{F}{x}{\alpha}$ and 
$H^* := \peval{H}{x}{\alpha}$.
Consider an arbitrary inequality $A$ in $F$.
In $H$, in the same position as $A$ appears in $F$, appears
$H_0 = \text{translate}(F,A)$.  Appearing at the same position in
$F^*$ is $A^* := \peval{A}{x}{\alpha}$, and at the same position in
$H^*$ is $H_0^* := \peval{H_0}{x}{\alpha}$.
If $A^*$ is a boolean constant, then $A^* = H_0^*$ by point (b) of
the specification of $\text{translate}(F,A)$.  If $A^*$ is $U_i$, for
some $i$, then $H_0^*$ is $\false$ by point (a) of the
specification of $\text{translate}(F,A)$.  Note that $A$ appears
unnegated in $F$ by the input requirements of translate$(\cdot)$.
If $A^*$ is neither a boolean constant nor a $U_i$, then it is an
inequality, possibly still with $\mydiv{\cdot}{\cdot}$ terms.
From line 21 of $\text{translate}(F,A)$, we see that
$H_0 = \neg G_1 \wedge H_1$, so
$H_0^* = \peval{\neg G_1 \wedge H_1}{x}{\alpha}
 = \neg\peval{G_1}{x}{\alpha} \wedge \peval{H_1}{x}{\alpha}$.
 $G_1$ only contains the free variables, so
 $\peval{G_1}{x}{\alpha}$ must be either $\true$ or $\false$.
 If it were $\true$, $H_0^*$ would be $\false$ which violates
 the specification of $\text{translate}(F,A)$, since $A^*$ is not also
 $\false$.  Thus, $\peval{G_1}{x}{\alpha} = \false$ so
 $H_0^* = \peval{H_1}{x}{\alpha}$.
 $\peval{H_1}{x}{\alpha}$ satisfies both output requirements of
 $\text{translate}({F^*},{A^*})$, and so
 $\peval{H_1}{x}{\alpha}$ is equivalent to $\text{translate}({F^*},{A^*})$.

 Let $\overline{F^*}$ be the result of replacing in $F^*$ each $U$
 variable with the constant $\false$. Since each $U$-variable appears
 unnegated in $F^*$, Theorem~\ref{theorem:UsVs2BoolConsts} guarantees
 that $\overline{F^*}$ and $F^*$ are fair-SAT equivalent.
 Let $\overline{H^*} :=  \text{translate}(\overline{F^*})$
 The quantifier blocks and boolean structures of $H^*$ and
 $\overline{H^*}$ are identical, with identical boolean constaints,
 and, at every position containing an inequality, equivalent
 inequalities.  By induction, $\overline{F^*}$ is fair-SAT if and only
 if $\overline{H^*}$ is $\true$.  Thus we have:
 $F^*$ is fair-SAT iff $\overline{F^*}$ is fair-SAT  iff $\overline{H^*}$ is $\true$ iff $H^*$ is $\true$.

 We next consider the case in which there are no free variables and
 the outermost quantifier block is existential. In this case, $F$ is
 fair-SAT if and only if there is an assignment of values to the
 variables of the outermost block for which the partial evaluation of
 $F$ is fair-SAT.  We can dispense with this case using an argument
 identical to the free variable case handled previously.

 Finally, we consider the case in which there are no free variables and
 the outermost quantifier block is universal. In this case, $F$ is
 fair-SAT if and only if $\flop{F}$ is not fair-SAT.
 By induction, $\flop{F}$ is fair-SAT if and only
 $\text{translate}(\flop{F})$ is true.
 By Lemma~\ref{lemma:flop2neg},
 $\text{translate}(\flop{F}) \Leftrightarrow \neg\text{translate}(F)$.
 So ...\\
 $F$ is fair-SAT
 $\Longleftrightarrow$
 $\flop{F}$ is not fair-SAT
 $\Longleftrightarrow$
 $\text{translate}(\flop{F})$ is $\false$
 $\Longleftrightarrow$
 $\text{translate}(F)$ is $\true$.
 \qed

\section{Example in {\sc Tarski}}
The translate algorithm described in this paper has been implemented
in the \tarski{} system \cite{BrownVale:2018,TarskiRepo2024} (version
1.41 and beyond).  We
include sample computations in Figure~\ref{figure:tarskisession}.
\begin{figure}
\begin{minted}[fontsize=\small]{nestedtext}
> (def F [all a, b[ex x[ b^2 + 4 a < 0 \/ x = 1/(a x + b)]]]) ; Example 10
:void
> F
[all a,b[ex x[b^2+4*a < 0 \/ x = 1/(a*x+b)]]]:uif
> (clear F 'noguard)
all a,b[ex x[b^2 + 4 a < 0 \/ a x^2 + b x - 1 = 0]]:tar
> (clear F 'naive)
all a,b[ex x[b^2 + 4 a < 0 \/ [a x^2 + b x - 1 = 0 /\ a x + b /= 0]]]:tar
> (clear F 'fair)
all a,b[ex x[b^2 + 4 a < 0 \/ [b = 0 /\ a = 0] \/ [a x^2 + b x - 1 = 0 /\ a x + b /= 0]]]:tar
> (qepcad-qe (clear F 'noguard))
false:tar
> (qepcad-qe (clear F 'naive))
false:tar
> (qepcad-qe (clear F 'fair))
true:tar
> ; Section 1.5: Under existing CA practice, formula reduces to false, as does its negation!
> (qepcad-qe (clear [ex x[ 1/x^2 < 0 ]] 'naive))
false:tar                                      
> (qepcad-qe (clear [~[ex x[ 1/x^2 < 0 ]]] 'naive))
false:tar
> ; With "fair-SAT" translation from paper, formula reduces to false, its negation to true.
> (qepcad-qe (clear [ex x[ 1/x^2 < 0 ]] 'fair))
false:tar                                     
> (qepcad-qe (clear [~[ex x[ 1/x^2 < 0 ]]] 'fair))
true:tar                                    
\end{minted}
  \caption{\tarski{} session using the formula from
    Example~\ref{ex:nullmatters} and example input from Section~\ref{section:capractice}}.
  \label{figure:tarskisession}
\end{figure}
Note that, prior to calling the
translation algorithm, the formula must be left ``uninterpreted'',
because some identities, like $\forall x[0 \cdot x = 0]$ cannot be
applied before the translation process.  This is why, in the example
\tarski{} session, the type annotation for the original input is
\verb|uif|, for ``uninterpreted formula'', as opposed to
\verb|tar|, for ``Tarski formula'', which appears after the
translation. 

The command \verb|(clear F pflag)| clears denominators in
uninterpreted formula \verb|F| following the process specified by
\verb|pflag|: \verb|'noguard| for clearing denominators without adding
guards;
\verb|'naive| for following the usual computer algebra system practice
of adding, for each denominator polynomial $p$, the guard $p \neq 0$
in conjunction with the constraint, cleared of denominators; and 
\verb|'fair| for the translate process defined in this paper.


\section{Conclusion}
This paper arose out of an attempt to produce a whitepaper to explain
why, for computer algebra solvers specializing in
quantified polynomial constraints, supporting division in a principled
way is impossible / impractical / undesirable --- that all we could do is require our clients,
whether human users or software systems interacting via APIs, to
provide us with ``well-defined'' input formulas, allowing us to clear
denominators naively and continue from that point with polynomials.
However, even that turned out to be difficult: what
constitutes a ``well-defined'' formula?

The project then morphed into trying to produce a notion of
``well-defined formula'' in the presence of divisions.  What should
one look for in a definition of a well-defined formula with divisions?
At a minimum, one would hope that a formula with divisions could be
made to be ``well defined'' by some form of guarding, which is a common
mathematical practice, and be invariant under some (many?) kinds of
rewritings, like moving around quantifiers and negations.
Moreover, it would be nice to have the
question of whether a formula is well defined be decidable.
Ultimately, this paper does indeed provide such a definition: a
formula is well-defined if the Tarski formula resulting from naively
clearing denominators (i.e. without guarding) is equivalent to the
Tarski formula produced by the \verb|translate| algorithm.  The
translate algorithm uses guarding, and the resulting definition of
``well-defined formula'' is invariant under many common rewritings.

However, what has resulted from this work is not just a definition of
``well-defined''.  Instead of showing why it is impossible (or at
least unreasonable) for a quantified polynomial constraint solver to
handle division in a principled way, the \verb|translate| algorithm
\emph{is} a principled way to handle division.  It has a well-defined
semantics and is not unduly sensitive to innocuous differences of
input formula phrasings. Moreover, if a user does not like those
semantics, for example does not like that $\forall x[1/x^2 \geq 0]$
evaluates to $\true$ under them, they have clear rules for what is
required to produce ``well-defined'' input formulas, whose meaning is
unaffected by the \verb|translate| algorithm.

It is important to also recognize the limitations to this approach of
handling division.  It does not make sense for theorem provers or SMT
solvers or other systems that manipulate formulas in first-order logic
generally, and for which real polynomial constraints represents just
one of many theories to be handled.  This is because with the fair-SAT
semantics, $\mydiv{\cdot}{\cdot}$ is not really a function in the
sense of first-order logic.  We see
this in the fact that 
$\exists x[x = 0 \wedge [1/x = 0 \vee \neg(1/x = 0)]]$
is not fair-SAT equivalent to 
$\exists x[x = 0 \wedge \true]$.
That said, if an SMT solver wants to use a computer algebra solver
as a ``theory solver'', this paper's definition of well-defined at least
provides a basis for how that communication should be handled in the
presence of divisions.

\bibliographystyle{plain}
\bibliography{mybib/BibEntries}

\end{document}